\documentclass[12pt]{article}
\usepackage[a4paper,left=2.5cm,right=2.5cm,top=2.5cm,bottom=2.3cm]{geometry}
\usepackage[utf8]{inputenc}
\usepackage{graphicx,subfigure}
\usepackage{epsfig}
\usepackage{mathrsfs}
\usepackage{amsmath}
\usepackage{amsfonts}
\usepackage{amssymb}
\usepackage{hyperref}
\allowdisplaybreaks
\usepackage[dvipsnames]{xcolor}

\newcommand{\be}{
\begin{eqnarray}
}

\newcommand{\beLabel}[1]{
\begin{eqnarray}
\label{eq:#1}
}

\newcommand{\nl}{
\nonumber \\
}

\newcommand{\flows}{
\tau
}

\newcommand{\ee}{
\end{eqnarray}
}

\newcommand{\eeLabel}{
\end{eqnarray}
}

\newcommand{\Lie}{\mathrm{Lie}}

\numberwithin{equation}{section}

\usepackage{fancyhdr}

\begin{document}

\title{{\bf Gradient flow of Einstein-Maxwell theory and Reissner-Nordstr\"om black holes} }
\author{Davide De Biasio,$^{\text{a}}$ Julian Freigang,$^{\text{a}}$ Dieter L\"ust$^{\text{a,b}}$ and Toby Wiseman$^{\text{c}}$ }
\date{}
\fancypagestyle{plain}{%
	\fancyhead[R]{LMU-ASC 30/22 \\ MPP-2022-130}
	\renewcommand{\headrulewidth}{0pt}
}
\maketitle

\begin{center}

\textit{$^{\text{a}}$
Max--Planck--Institut f\"ur Physik, Werner--Heisenberg--Institut,}\\ \textit{ F\"ohringer Ring 6, 80805 M\"unchen, Germany}
\vspace{0.5cm}

\textit{$^{\text{b}}$Arnold Sommerfeld Center for Theoretical Physics, }\\ \textit{Ludwig Maximilians Universit\"at M\"unchen,} \\ \textit{  Theresienstrasse 37, 80333 M\"unchen, Germany}
\vspace{0.5cm}

\textit{$^{\text{c}}$Theoretical Physics Group, Blackett Laboratory, Imperial College London}\\ \textit{London SW7 2AZ, UK}
\vspace{1cm}

{\small
\texttt{debiasio@mpp.mpg.de}, \texttt{J.Freigang@physik.uni-muenchen.de}, \texttt{dieter.luest@lmu.de}, 
\texttt{t.wiseman@imperial.ac.uk}
}

\end{center}

\vspace{1cm}
\begin{abstract} 

Ricci flow is a natural gradient flow of the Einstein-Hilbert action. Here we consider the analog for the Einstein-Maxwell action, which gives Ricci flow with a stress tensor contribution coupled to a Yang-Mills flow for the Maxwell field. We argue that this flow is well-posed for static spacetimes with pure electric or magnetic potentialsand show it preserves both non-extremal and extremal black hole horizons. In the latter case we find the flow of the near horizon geometry decouples from that of the exterior.
The Schwarzschild black hole is an unstable fixed point of Ricci flow for static spacetimes. 
Here we consider flows of the Reissner-Nordstr\"om (RN) fixed point. The magnetic RN solution becomes a stable fixed point of the flow for sufficient charge. 
However we find that the electric RN black hole is always unstable. 
Numerically solving the flow starting with a spherically symmetric perturbation of a non-extremal RN solution, we find similar behaviour in the electric case to the Ricci flows of perturbed Schwarzschild, namely the horizon shrinks to a singularity in finite time or expands forever. In the magnetic case, a perturbed unstable RN solution has a similar expanding behaviour, but a perturbation that decreases the horizon size flows to a stable black hole solution rather than a singularity.
For extremal RN we solve the near horizon flow for spherical symmetry exactly, and see in the electric case two unstable directions which flow to singularities in finite flow time. However, even turning these off, and fixing the near horizon geometry to be that of RN, we numerically show that the flows appear to become singular in the vicinity of its horizon.

\end{abstract}

\newpage

\setcounter{page}{1}

\tableofcontents

\section{Introduction}
\label{sec:intro}
Ricci flow~\cite{hamilton1982three,Friedan:1980jm} may be regarded as a natural gradient flow of the Euclidean signature Einstein-Hilbert action with respect to the DeWitt metric~\cite{Headrick:2006ti,Tseytlin:2006ak}, and provides a way to study both the stability of saddle points in the canonical ensemble, as well as the global structure of the space of solutions. An example is gravity in a box with spherical spatial symmetry and fixed boundary radius. Describing this at finite temperature using a Euclidean continuation, the saddle points of the action are the small and large black holes, as well as `hot' flat space, and  being Ricci flat, these are fixed points of Ricci flow. The small black hole is always thermodynamically unstable in the canonical ensemble~\cite{York:1986it} and there is an analog of the Hawking-Page phase transition between the large black hole and hot flat space~\cite{Hawking:1982dh}. This thermodynamic instability of the small black hole results in a negative mode of the Lichnerowicz operator~\cite{Gross:1982cv,Whiting:1988qr,Whiting:1988ge,Prestidge:1999uq,Reall:2001ag}, which drives an instablity under Ricci flow~\cite{Headrick:2006ti}.
A very interesting recent study of the Wick rotation from Lorentzian signature find that mode stability about a fixed point in the semiclassical Euclidean path integral is directly related to stability under Ricci flow~\cite{Marolf:2022ntb}.

This instability implies a relevant deformation of the small black hole that generates two flows away from it. In~\cite{Headrick:2006ti} it was found these flows connect to the other two stable fixed points, the large black hole, and hot flat space, the latter requiring a surgery to resolve a singularity along the flow. 
This is intruigingly qualitatively similar to the behaviour of real time evolution of a small black hole surrounded by a bath of radiation at its Hawking temperature.
Removing the box by taking it to infinite size, the small black hole tends to the usual Schwarzschild solution and the large black hole is removed with the box. Since the unstable mode is localized about the horizon, the instability under Ricci flow persists. One of the two flows from Schwarzschild then again tends to flat spacetime (after resolving a singularity at finite flow time that changes the spacetime topology). The other `eats up' spacetime indefinitely as there is now no large black hole for it to flow to.

These Ricci flows of the Schwarzschild black hole can be thought of as RG flow of a sigma model~\cite{Friedan:1980jm}.
However it is also possible that the off-shell geometries that the flow passes through may have significance in the gravitational path integral~\cite{Headrick:2006ti}.
In the context of string theory phenomenology, Ricci flow has also been suggested as a tool to refine and extend the so-called Swampland distance conjecture \cite{Ooguri:2006in,Kehagias:2019akr}. In particular, pushing the intuition behind the AdS conjecture \cite{lust2019ads} even further, it allows a generalization of infinite distance moduli space limits (which are expected to be accompanied by infinite towers of light states) to the case of displacements in the metric components.

The above discussion was for Euclidean signature. In fact Ricci flow is not well defined for Lorentzian spacetimes in general. However, restricting to static or stationary metrics it is well posed and preserves smooth non-extremal Killing horizons~\cite{Headrick:2006ti,Adam:2011dn,Wiseman:2011by}, the surface gravity and angular velocities of these being constant along the flow. 
Furthermore, the Euclidean negative mode discussed above preserves the Euclidean $U(1)$ isometry, and hence can be Wick rotated to an instability of the Ricci flow of static black hole spacetimes. Thus the unstable flows of Euclidean Schwarzschild can be Wick rotated to give flows of static spacetimes away from Lorentzian Schwarzschild.

Motivated by these considerations it is interesting to consider flow of a gravitational system with matter. 
Here we will consider the simple setting of gravity coupled to a Maxwell field. 
There is a natural family of flows that are gradient flows of the Einstein-Maxwell action with respect to a natural metric on the superspace. These flows couple the metric and Maxwell vector potential.
The flow of the metric is similar to Ricci flow but with an extra contribution from the Maxwell stress tensor. The flow of the Maxwell field is given by the Yang-Mills flow in the background of the metric. We call this combined gradient flow the Einstein-Maxwell flow (EM flow). The non-trivial fixed points of the flow are then the Reissner-Nordstr\"om (RN) solutions.

Firstly we introduce the flow in section~\ref{sec:EMFlow}, showing that when it is restricted to static metrics with either electric or magnetic potentials, then it is well-posed in the sense that it can be made parabolic in character.
We emphasize that general potentials with both electric and magnetic parts give rise to Poynting energy flux, and their backreaction is therefore not compatible with static symmetry even though the electric and magnetic fields are static (see~\cite{FeynLect} for a discussion on this).
Hence here, with the restriction to static metrics, we will only consider purely electric or purely magnetic Maxwell fields.
In section~\ref{sec:bhs} we show that this flow of static metrics and Maxwell potentials preserves both non-extremal and extremal horizons.
\footnote{
In~\cite{Figueras:2011va} Ricci flow was discussed for geometries with extremal horzions where the near horizon geometry was a solution of the Einstein equations, so was fixed in flow time and consequently could be viewed as a Dirichlet boundary condition. Here we consider the more general case where the horizon geometry flows as well.
} 
We find a new feature of the flows of spacetimes with extremal horizons is that, as one might expect, the flow of the near horizon geometry decouples from the flow of the exterior spacetime.
\footnote{
For an excellent review of extremal horizons and their near horizon geometry see~\cite{Kunduri:2013gce}.}
Both flows preserve the surface gravity of the solution in the non-extremal setting. 
We show that the electric flow preserves the electric potential difference, but argue it does not preserve the electric charge. Conversely we show that the magnetic flow does preserve the magnetic charge. 

In section~\ref{sec:RNflows} we consider the stability of the RN fixed point for non-extremal solutions in the spherically symmetric sector using numerical methods both for the case of  electric and  magnetic charge. Negative modes of the (non-extremal) Euclidean magnetic RN solution were studied by Monteiro and Santos~\cite{Monteiro:2008wr}. They considered the negative mode of Reissner-Nordstr\"om and found that precisely where the specific heat at fixed (magnetic) charge becomes positive, the negative mode disappears. These Euclidean negative modes can be Wick rotated to give an unstable mode of the static magnetic EM flow we consider here. The expectation is then that RN is a stable fixed point of the magnetic EM flow for sufficient magnetic charge, and this is indeed what we see. 
We then consider the non-extremal RN fixed point of the electric EM flow, which we recall preserves the electric potential difference. Now an interesting fact is that in the
grand canonical ensemble, where we fix temperature and electric potential, the RN solution has an unstable thermodynamics up to extremality. One might then think this suggests that RN will be an unstable fixed point of the electric EM flow all the way up to extremal electric charge. However due to there being no simple Wick rotation to Euclidean solutions in this electric setting, we know of no formal correspondence between the static flow stability, Euclidean stability or thermodynamics. Nonetheless our numerical analysis of the linear behaviour of the electric EM flow about RN indicates a complicated structure of unstable modes which indeed appear to persist to the extremal limit. In particular we find more unstable modes as charge increases, rather than less as in the magnetic case. 

Having discussed the linear stability of non-extremal RN for the electric and magnetic flows we then discuss where the unstable modes of these non-extremal RN solutions flow to by simulating the spherically symmetric non-linear flow equations numerically. 
In the electric case, for all electric charge the RN fixed point is unstable and we find qualitatively similar flow behaviour as for the uncharged Schwarzschild fixed point under Ricci flow. Perturbing the unstable electric RN fixed point to initially reduce the horizon size results in a flow that shrinks the horizon to a zero size singularity in finite flow time, whereas initially expanding the horizon gives a flow that `eats up' spacetime, apparently indefinitely.
One might expect similar behaviour to the uncharged case, as the flow preserves the surface gravity and electric potential, and with these fixed values there is only one infilling RN solution. 
In the magnetic case, for large charge the RN solution is linearly stable under the magnetic EM flow, and we find large perturbations flow back to it. However for the small charge case, the RN fixed point is unstable, there also exists a stable fixed point with smaller horizon size and the same magnetic charge and surface gravity (which are both preserved by the magnetic EM flow). We find that perturbing the unstable fixed point so that the horizon initially decreases in size, gives a flow that asymptotes to this smaller stable solution. An initial perturbation that expands the horizon results in a flow that again 'eats up' spacetime, as in the electric and uncharged cases.

Finally in section~\ref{sec:ExtRNflows} we study the extremal RN fixed points, and here we are able to analytically solve the full non-linear near horizon flow for spherical symmetry. As expected, we find the magnetic extremal near horizon geometry is stable, and the electric one is unstable, and we elucidate the various singularities that these flows tend towards. Again restricting to spherical symmetry, we perform numerical simulation of the flow of the full extremal black hole spacetimes, and again find stability in the magnetic case, where initial configurations that are non-linear deformations of extremal RN, preserving extremality, flow back to the fixed point. 
In the electric case if the near horizon geometry is perturbed our near horizon analysis already implies the spacetime flows away from RN. However one might not perturb the near horizon geometry, which is then fixed under the flow, and then wonder whether extremal RN is stable under deforming the exterior spacetime. Our simulations show that it is still unstable, and furthermore we find that a curvature singularity appears to develop just outside the horizon under this flow.

\section{Einstein-Maxwell theory}
\label{sec:setup}

This paper will be concerned with a natural gradient geometric flow of the static Einstein-Maxwell system. Having derived the flow and discussed its well-posedness, we then focus on its natural fixed points, the Reissner-Nordstr\"om (RN) solutions, both non-extremal and extremal, with a view to determining their stability.

Therefore to begin our discussion we will briefly review the 4-dimensional Einstein-Maxwell system and its RN black hole solutions. We will also introduce convenient coordinates for the later discussion. The theory contains the metric $g_{\mu\nu}$ and gauge field $A_\mu$ and classical solutions solve the Einstein-Maxwell equations, \be
R_{\mu\nu} = 2 F_{\mu\alpha} F_\nu^{~~\alpha} - \frac{1}{2} F^2 g_{\mu\nu} \; , \quad \nabla^\mu F_{\mu\nu} = 0
\ee
which derive from the action,
\be
S = \frac{1}{16 \pi G} \int d^4x \sqrt{-g} \left( R - F^2 \right) \, .
\ee
We note that we have chosen to use a gravitational normalization for the Maxwell field.
The static vacuum solutions to this are the Reissner-Nordstr\"om black holes. The metric may be written as,
\be
ds^2 = - f(r) dt^2 +   \frac{dr^2}{f(r)}  + r^2 d\Omega^2 \; , \quad  f(r) = \left( 1 - \frac{r_+}{r} \right)  \left( 1 - \frac{r_-}{r} \right) 
\ee
with $d\Omega^2 = d\theta^2 + \sin^2{\theta} d\phi^2$, and the magnetically and electrically charged solutions have gauge potentials,
\be
\label{eq:RNgaugemag}
A_{mag} = - \sqrt{r_+ r_-} \cos{\theta} d\phi \; , \quad A_{elec} =  \sqrt{r_+ r_-} \left( \frac{1}{r_+} - \frac{1}{r} \right) dt
\ee
respectively.
\footnote{
One may also source the RN solution with a mixed electromagnetic solution of the form $A = \alpha \left(  \frac{1}{r_+} - \frac{1}{r} \right) dt + \beta \cos{\theta} d\phi$ for appropriate $\alpha$, $\beta$. Here, as stated earlier, we will only be concerned with purely electric or magnetic solutions as generally mixed fields will produce a Poynting energy flux whose backreaction is not compatible with the static symmetry of the metric. In the case of these spherical solutions a static solution with both electric and magnetic charge is allowed since the spherical symmetry ensures the electric and magnetic fields are parallel, being radially directed, and hence the Poynting vector vanishes.
}
The mass $M$ and magnetic or electric charge $Q$ (in appropriate units) of the solutions are,
\be
2 G M =  r_+ + r_-  \; , \quad  Q = \sqrt{r_+ r_-} 
\ee
so that $G M \ge | Q |$. We may also consider the negatively charged solutions by reversing the signs of the gauge potentials above, but here just consider the positively charged case for simplicity.
Regarding $r_+$ as the outer horizon position implies $r_+ > r_-$.

The solutions obey a First Law, $dM = T d S + \mu d Q$, where the Hawking temperature, $T$, Bekenstein-Hawking entropy $S$, and potential $\mu$ associated to the charge, $Q$, of the non-extremal solutions are given by,
\be
T = \frac{\kappa}{2\pi} = \frac{r_+ - r_-}{4\pi r_+^2} \; , \quad S = \frac{ \pi r_+^2}{G} \; , \quad \mu = \frac{1}{G} \sqrt{\frac{r_-}{r_+}} 
\ee

where $\kappa$ is the surface gravity.
We note that for the electric RN, one can see from writing $A_{elec} = -\psi \,dt$ that $G \mu$ corresponds to the potential difference from the horizon to that at infinity, so $\psi_{r_+} - \psi_{\infty}$, noting that the potential at the horizon vanishes.
The extremal limit corresponds to taking $r_- \to r_+$, so that the surface gravity vanishes, and corresponds to taking the charge $| Q | \to G M$. 
The specific heat capacity at constant charge, $c_Q$, and capacitance at fixed temperature, $\epsilon_T$, are;
\be
c_Q = T \left. \frac{\partial S}{\partial T} \right|_Q =  - \frac{2 \pi r_+^2}{G} \frac{ \left( r_+ - r_- \right)}{\left( r_+ - 3 r_- \right) }  \; , \quad \epsilon_T = \left. \frac{\partial Q}{\partial \mu} \right|_T = \frac{G r_+ ( r_+ - 3 r_- )}{r_+ - r_-}
\ee
respectively. We note that $c_Q$ is negative for small charges, and becomes positive for $r_+  \ge r_- > \frac{1}{3} r_+$, so for charges $| Q | > \frac{\sqrt{3} G M}{2} = Q_{crit}$, and that $\epsilon_T$ has precisely the opposite sign to $c_Q$.
If we work at fixed temperature $T$ and charge $Q$, the relevant ensemble is the canonical ensemble, and as discussed in~\cite{Monteiro:2008wr,Monteiro:2009tc} then thermodynamic stability is simply governed by positivity of $c_Q$. Thus at fixed temperature and charge, the canonical ensemble is unstable for $| Q | < Q_{crit}$, but becomes stable for sufficient charges $| Q | > Q_{crit}$.
If we instead choose to work at fixed temperature $T$ and potential $\mu$, then we must use the grand canonical ensemble, and the condition for stability, namely that the corresponding Gibbs potential has a local minimum, is then given by positivity of \emph{both} $c_Q$ as well as $\epsilon_T$ (see for example~\cite{LandauLifshitz}). Now we see from their explicit forms that $c_Q$ and $\epsilon_T$ have opposite sign, and hence the grand canonical ensemble is unstable for all values of temperature and potential.

The work of~\cite{Prestidge:1999uq,Reall:2001ag} related the negative mode of the Euclidean continuation of Schwarzschild to its negative specific heat capacity, and hence thermodynamic instability. The magnetic RN solution can also be analytically continued to a smooth Euclidean solution in the same manner. Monteiro and Santos~\cite{Monteiro:2008wr} have shown it also has a Euclidean negative mode for charges where the specific heat at constant charge is negative, and hence the canonical ensemble is unstable, and this mode disappears as the charge is increased to $|Q |= Q_{crit}$ (see also~\cite{Prestidge:PhD}). 
However for electric RN, one fixes $A_{elec}$ at infinity, and therefore this is naturally related to the ensemble with fixed electric potential. This is therefore associated to the unstable grand canonical ensemble. However the static Lorentzian electric RN solution cannot straightforwardly be analytically continued to a (real) Euclidean solution, due to the electric gauge field whose potential would naively become imaginary. Interestingly, however, we will find later that the stability of the static EM flows about RN do reflect these behaviours in the sense that the flow of the magnetically charged fixed point becomes stable for sufficient charge, but that of the electric fixed point does not.
This is particularly interesting in light of the recent links between the stability of the canonical ensemble for Schwarzschild in a box and its stability under Ricci flow~\cite{Marolf:2022ntb}, and suggests a generalization which includes charge.

We will find our geometric flows preserve surface gravity of a horizon, and hence we will be interested in the set of solutions at a fixed temperature in order to understand possible end-points of the flow. Furthermore, we will see that the magnetic flow preserves magnetic charge, and the electric flow preserves the electric potential difference between the horizon and infinity. From the expressions above it follows that for magnetic RN solutions, at a given  magnetic charge there is a maximum temperature,
\be
T_{max} = \frac{1}{6 \sqrt{3} \pi |Q|} \; .
\ee
This arises since for fixed charge there is a minimum mass, given by the BPS bound $M \ge |Q|$, and a unique solution exists for all greater masses. Then for $M \to |Q|$ the temperature vanishes, but for very large $M \gg |Q|$ the temperature also becomes small, so by continuity in between there is a maximum. Assuming one takes a temperature $T < T_{max}$ there are then two magnetic RN solutions. In analogy with Hawking-Page we may think of these as a small and large black hole. Now the smaller one, with mass $M < \frac{2 |Q|}{\sqrt{3} G}$, is thermodynamically stable (with no Euclidean negative mode), and the larger one has a greater horizon size and mass $M > \frac{2 |Q|}{\sqrt{3} G}$, and is thermodynamically unstable (with a Euclidean negative mode). Given a charge, then increasing the temperature to its critical value, $T_{max}$, these two solutions merge to a single marginally stable solution. For greater temperatures there are no RN black hole spacetimes.
For electric RN solutions, then fixing the electric potential the situation is different. For a given potential there exists a single black hole for every temperature, and these solutions are always thermodynamically unstable.

When considering a RN fixed point of the flow, we will find it convenient to choose units so that $r_+ = 1$, so that the outer horizon is a unit sphere, and the exterior spacetime is $r > 1$. In this case, a useful radial coordinate we will use later is $\rho$, related to $r$ by,
\be
r = \frac{1}{1-\rho^2}
\ee
which compactifes the infinite range $r \in [1, \infty)$ to the interval $\rho \in [ 0, 1 )$. 
%
%
Then the RN metric (in these units, so $r_+ = 1$) becomes,
\be
\label{eq:RNmetric}
ds^2 = - \rho^2 F(\rho) dt^2 + 4 r(\rho)^4 \frac{d\rho^2}{F(\rho)}  + r(\rho)^2 d\Omega^2 \; , \quad  F(\rho) = 1 - r_- + r_- \rho^2 
\ee
with the magnetic Maxwell potential as above, and the electric one being,
\be
\label{eq:RNgaugeelec}
A_{elec} = \sqrt{r_-} \rho^2 dt \, .
\ee
Taking $r_- \to 0$ yields the uncharged Schwarzschild solution, whereas taking $r_- \to 1$ gives the extremal RN black holes. In this extremal limit we introduce a new coordinate,
\be
\rho' = \rho^2
\ee
so that again $\rho' \in [ 0, 1 )$, and then
 the extremal metric is,
\be
\label{eq:RNextmetric}
ds^2 = - \rho'^2  dt^2 +  r(\rho')^4 \frac{d\rho'^2}{\rho'^2}  + r(\rho')^2 d\Omega^2 \; , \quad  r = \frac{1}{1-\rho'}
\ee
now with either an electric or magnetic potential,
\be
A_{elec} =  \rho' dt 
\; , \quad
A_{mag} =-  \cos{\theta} d\phi 
\ee
for positively charged solutions.
Taking $\rho' \to 0$, so that $r \to 1$, we recognise the near horizon geometry as $AdS_2 \times S^2$.

\section{Gradient flow of the Einstein-Maxwell action}
\label{sec:EMFlow}

Given an action $S$ which is a functional of some fields $\Psi^A$, then the variation $\delta S/\delta \Psi^A$ is a covector on the space of fields, the `superspace'. As discussed in~\cite{Headrick:2006ti}, then given a metric on superspace, $G_{AB}$, which is a functional of the fields $\Psi^A$, this may be converted to a vector, and a gradient flow is defined as,
\be
\label{eq:gradflow}
\frac{d}{d\lambda} \Psi^A = G^{AB} \frac{\delta S}{\delta \Psi^{B}} \; .
\ee
The simplest such local superspace metrics will only depend on the fields and not their derivatives.
If $G_{AB}$ is a positive definite metric, then this is a gradient descent of the action, which monotonically decreases along the flow. Different choices of superspace metric will yield different flow equations.

Let us consider the Einstein-Hilbert action in this context,
\be
S_{EH} = \frac{1}{16 \pi G} \int d^4x \sqrt{-g} R
\ee

We may associate $\Psi^A$ above as the metric $g_{\mu\nu}(x)$, and the abstract index $A$ as enumerating both the metric components and coordinate position.
A natural metric on superspace is given by the DeWitt${}_k$ metric (using the notation of~\cite{Marolf:2022ntb}),
\be
( h_{\mu\nu}, \tilde{h}_{\alpha\beta} ) =  \frac{1}{32 \pi G} \int d^4x \sqrt{-g} h_{\mu\nu} G^{\mu\nu\alpha\beta} \tilde{h}_{\alpha\beta}
\ee
where $h_{\mu\nu}$ and $\tilde{h}_{\alpha\beta}$ are two perturbations of the metric $g_{\mu\nu}$, and,
\be
G^{\mu\nu\alpha\beta} = \frac{1}{2} \left( g^{\mu\alpha} g^{\nu\beta} + g^{\mu\beta} g^{\nu\alpha} + k g^{\mu\nu} g^{\alpha\beta} \right)
\ee
for real $k$.
Then rearranging the gradient flow equation~\eqref{eq:gradflow} as, $G_{AB} ( d\Psi^A/d\lambda  ) \delta \Psi^{B} = \delta S$, with $\delta S$ given by the variation $\delta \Psi^B$, explicitly gives,
\be
\left( \frac{d}{d\lambda} g_{\mu\nu}, \delta g_{\alpha\beta} \right) = \delta S_{EH} = - \frac{1}{16 \pi G} \int d^4x \sqrt{-g} \delta g_{\alpha\beta} \left( R^{\alpha\beta} - \frac{1}{2} g^{\alpha\beta} R \right) \; .
\ee
We will focus on the case $k = -1$, and then this gives the flow equation as,
\be
\frac{d}{d\lambda} g_{\mu\nu}(x) =  - 2 R_{\mu\nu}
\ee

which is simply the Ricci flow. An important point is that since the DeWitt${}_{-1}$ metric is not positive definite, the Einstein-Hilbert action does not generally vary monotonically along Ricci flows. A monotone functional is famously given by the Perelmann entropy~\cite{Perelman:2006un}.

Fixed points of Ricci flow are ones so that $\dot{g}_{\mu\nu}  = 0$ where $\dot{} = d/d\lambda$, and are thus Ricci flat metrics. However one may also consider solutions to the Ricci soliton equation,
\be
R_{\mu\nu} = \nabla_{(\mu} \xi_{\nu)}
\ee
to be geometric fixed points of the flow, since for these we have $\dot{g}_{\mu\nu} = - \Lie_\xi g_{\mu\nu}$. For non-vanishing $\xi$ the metric is clearly varying, but the geometry it represents is not. Thus a Ricci soliton represents a geometric fixed point of the flow, so the geometry doesn't flow, but the coordinates it is presented in do.

Ricci flow is itself diffeomorphism invariant. 
By performing a flow time dependent diffeomorphism,
\be
x^\mu \to x^\mu + v^\mu(\lambda,x) \; , \quad \frac{d}{d\lambda} v^\mu = \xi^\mu
\ee
then we generate a new flow of the metric,
\be
\frac{d}{d\lambda} g_{\mu\nu} = - 2 R_{\mu\nu} + \mathrm{Lie}_\xi g_{\mu\nu}
\ee
where we may write $\Lie_\xi g_{\mu\nu} = 2 \nabla_{(\mu} \xi_{\nu)}$.
The flows $g_{\mu\nu}(\lambda)$ for the Ricci flow and the above flow are the same geometrically, but the two metrics will differ by a coordinate transformation at any given $\lambda$, this transformation depending on the flow time.

Due to its diffeomorphism invariance, Ricci flow is not well-posed as a p.d.e. for initial data at some starting flow time $\lambda$. However, following DeTurck~\cite{DeTurck1983DeformingMI}, we may make the choice that,
\be
\label{eq:DeTurckxi}
\xi^\mu = g^{\alpha\beta} \left( \Gamma^\mu_{~~\alpha\beta} - \bar{\Gamma}^\mu_{~~\alpha\beta} \right)
\ee 
given a smooth fixed (i.e. flow time independent) reference connection $\bar{\Gamma}^\mu_{~~\alpha\beta}$ on the manifold. Then the principle part of the flow is,
\be
\frac{d}{d\lambda} g_{\mu\nu} =_{PP} g^{\alpha\beta} \partial_\alpha \partial_\beta g_{\mu\nu}
\ee
and for Riemannian metrics this is parabolic. For Lorentzian signature metrics this is generally not well-posed, as timelike perturbations `anti-diffuse'. However following~\cite{Adam:2011dn,Wiseman:2011by} we may restrict to the space of static or stationary metrics, and then the flow is indeed parabolic, and thus well-posed.

Let us take the reference connection $\bar{\Gamma}^\mu_{~~\alpha\beta}$ to be the Levi-Civita connection of a static reference metric, $\bar{g}_{\mu\nu}$. Then the flow  obviously preserves staticity of the metric -- starting with a static metric, it will remain static. If the metric is static we may generally write it locally as,
\be
g_{\mu\nu}(x) = \left( 
\begin{array}{cc}
- N(x^k) & 0 \\ 0 & h_{ij}(x^k)
\end{array}
\right)
\ee
with $N \ge 0$ and $\det(h_{ij}) > 0$, using coordinates $x^\mu = (t, x^k)$ adapted to the static symmetry. Note that in these static coordinates $N$ may vanish at black hole horizons, but away from these it should be positive. We may always choose coordinates locally so that the spatial metric $h_{ij}$ is a smooth Riemannian metric.
Then since $g_{\mu\nu}$ has no explicit time dependence, the principle part becomes,
\be
\frac{d}{d\lambda} g_{\mu\nu} =_{PP} h^{ij} \partial_i \partial_j g_{\mu\nu} 
\ee
which is indeed a parabolic diffusion-like flow for the metric components on the curved space $h_{ij}$.

We now consider gradient flow of the Einstein-Maxwell action,
\be
S = \frac{1}{16 \pi G} \int d^4x \sqrt{- g} \left( R - F^2 \right)
\ee
with $F_{\mu\nu} = 2 \partial_{[\mu} A_{\nu]}$. Now schematically our field space $\Psi^A$ is composed of the metric $g_{\mu\nu}(x)$ together with the gauge field $A_\mu(x)$.
A natural superspace metric is then given by
\be
( h_{\mu\nu}, \tilde{h}_{\alpha\beta} ) =  \frac{1}{32 \pi G} \int d^4x \sqrt{-g} h_{\mu\nu} G_{(g)}^{\mu\nu\alpha\beta} \tilde{h}_{\alpha\beta} \; , \quad( u_{\mu}, \tilde{u}_{\alpha} ) =  \frac{1}{16 \pi G} \int d^4x \sqrt{-g} u_{\mu} G_{(A)}^{\mu\alpha} \tilde{u}_{\alpha} \nl
\ee
where, as above, $h_{\mu\nu}$, $\tilde{h}_{\alpha\beta}$ are metric perturbations, and $u_{\mu}$, $\tilde{u}_{\alpha}$ are perturbations to the gauge field $A_\mu$, and we take,
\be
G_{(A)}^{\mu\alpha} = \frac{4}{\flows} g^{\mu\alpha} \; , \quad
G_{(g)}^{\mu\nu\alpha\beta} = \frac{1}{2} \left( g^{\mu\alpha} g^{\nu\beta} + g^{\mu\beta} g^{\nu\alpha} - g^{\mu\nu} g^{\alpha\beta} \right) 
\ee
for a constant $\flows$, and further we take $( h_{\mu\nu}, u_\alpha ) = 0$ so that the metric and gauge perturbations are orthogonal. We may regard the parameter $\flows$ as determining the speed of the flow of the gauge field relative to that of the metric, and later we will focus on the case $\flows = 1$.
We see that this superspace metric is constructed only from the metric (so doesn't involved the gauge field) and that its restriction to the metric is the DeWitt${}_{-1}$ superspace metric.
Then
we obtain the gradient flow equation,
\be
\label{eq:EMflowNoDT}
\frac{d}{d\lambda} g_{\mu\nu} = - 2 R_{\mu\nu} + 4 F_{\mu\alpha} F_\nu^{~\alpha} -  F^2 g_{\mu\nu} \; , \quad \frac{d}{d\lambda} A_\mu = \flows \nabla^\alpha F_{\alpha\mu} \; .
\ee

We will call this the Einstein-Maxwell (or EM) flow, and its fixed points, $\dot{g}_{\mu\nu} = 0 = \dot{A}_\mu$, are solutions of the Einstein-Maxwell equations. 
This flow is both diffeomorphism and gauge invariant.
Thus we may consider the analog of Ricci solitons, so field configurations where the geometry and gauge field flow only up to diffeomorphisms and gauge transformations,
\be
  \dot{g}_{\mu\nu} = - \Lie_\xi g_{\mu\nu} \; , \quad  \dot{A}_{\mu} = - \Lie_\xi A_{\mu} - \partial_\mu \Lambda 
\ee
with $\xi$ generating the flow dependent diffeomorphism, and $\Lambda$ the gauge transformation of the Maxwell field along the flow.
We may explicitly modify the flow to add such a flow time dependent diffeomorphism $\xi$ and gauge transformation $\Lambda$, as,
\be
\label{eq:EMflow}
\frac{d}{d\lambda} g_{\mu\nu} =  \Lie_\xi g_{\mu\nu} - 2 R_{\mu\nu} + 4 F_{\mu\alpha} F_\nu^{~\alpha} - F^2 g_{\mu\nu} \; , \; \frac{d}{d\lambda} A_\mu = \Lie_\xi A_{\mu} + \partial_\mu \Lambda + \flows \nabla^\alpha F_{\alpha\mu} 
\ee
to give the same flow, up to diffeomorphisms and gauge transformations. 
The analog of Ricci solitons can then be thought of as the fixed points of this flow.

As for Ricci flow, the Einstein-Maxwell flow is not parabolic, even when restricted to static field configurations. In a linearization of the flow, the two derivative terms vanish on linear perturbations that are diffeomorphisms, or gauge transformations of the gauge field. In order to obtain a well-posed flow, we take the modified flow above in equation~\eqref{eq:EMflow} and choose $\xi$ as the DeTurck vector in equation~\eqref{eq:DeTurckxi}, and further take,
\be
\label{eq:gauge}
 \Lambda = \flows \nabla^\alpha A_\alpha \, .
\ee
Then choosing the reference metric to be static, the flow truncates to the space of static metrics and gauge fields.
The principle part of the flow on static metrics and gauge fields then becomes,
\be
\frac{d}{d\lambda} g_{\mu\nu} =_{PP} h^{ij} \partial_i \partial_j g_{\mu\nu} \; , \quad \frac{d}{d\lambda} A_\mu =_{PP} \flows h^{ij} \partial_i \partial_j  A_\mu
\ee
and since $h_{ij}$ is a smooth Riemannian metric, the character of the flow is governed by the components of $g_{\mu\nu}$ and $A_{\mu}$ diffusing on the spatial geometry $h_{ij}$, and hence is parabolic assuming that we choose $\flows > 0$.

We have written this static flow for a general gauge potential $A_\mu$ compatible with the static symmetry.
However, as emphasized earlier, in general a static potential will have a stress tensor that is only stationary, but not static due to a Poynting energy flux being generated when both (static) electric and magnetic fields are present. Thus here we take the Maxwell potential to be either purely electric, $A = -\psi dt$ (where the gauge function $\Lambda = \flows \nabla^\alpha A_\alpha$ vanishes), or purely magnetic so $A = A_i dx^i$. These electric or magnetic forms are consistently preserved by the flow for the DeTurck choice of $\xi$ and the gauge choice in equation~\eqref{eq:gauge}.

From this point on, we will term the parabolic Einstein-Maxwell flow in~\eqref{eq:EMflow} of a static spacetime, with either magnetic or electric Maxwell field, and DeTurck vector $\xi_\mu$ and gauge choice $\Lambda$ as in equation~\eqref{eq:gauge}, as the ``magnetostatic or electrostatic Einstein-Maxwell (EM) flow''.

\section{Static EM flows of black hole spacetimes}
\label{sec:bhs}

We now consider the EM flow on static black hole spacetimes. While the flow is parabolic away from horizons, it isn't clear that it will preserve the smooth structure of a horizon, and this is what we aim to demonstrate. Further in the extremal case we will see the nice property that the flow of the near horizon geometry decouples from the flow of the exterior of the horizon as one might expect.

\subsection{Static non-extremal black hole spacetimes}
\label{sec:nonextremal}

The static EM flow of static non-extremal black holes is very simple to consider, being similar to the discussion for Ricci flow.  In fact for Ricci flow one may simply perform a Euclidean continuation of a static metric, identifying the period of the Euclidean time coordinate so that the Euclidean manifold becomes smooth at the horizon. Since Ricci flow of this Euclidean geometry preserves the $U(1)$ isometry associated to the static symmetry, then it is obvious that the smoothness of the horizon is preserved.

However in the case of the EM flow a gauge field with electric potential component cannot generally be analytically continued to Euclidean time. Hence we will use a different argument, that of~\cite{Adam:2011dn,Wiseman:2011by}, which exploits the similarity between a static Killing horizon and the origin of polar coordinates. Following~\cite{Adam:2011dn,Wiseman:2011by} the most general smooth static symmetric metric with a Killing horizon with surface gravity $\kappa$ (with respect to the Killing vector $\partial/\partial t$), can be written in coordinates $x^\mu = (t, r, x^a)$ adapted to the static symmetry and horizon as,
\be
ds^2 = - r^2 V dt^2 + U ( dr + r U_a dx^a )^2 + h_{ab} dx^a dx^b
\ee
where $r = 0$ is the horizon, and $V, U, U_a, h_{ab}$ are all smooth functions of $r^2$ and of $x^a$, with $V, U > 0$, and $h_{ab}$ is a Riemannian metric, and at the horizon there is the additional condition that,
\be
\label{eq:surfgravcondition}
\left. \left( V - \kappa^2 U \right) \right|_{r=0} = 0 \, .
\ee
Then making the coordinate transformation $\alpha = r \sinh \kappa t$, $\beta = r \cosh \kappa t$ (analgous to that going from polar to Cartesians coordinates) one finds a metric tensor with smooth components and non-vanishing determinant.

The most general symmetric two tensor that is also smooth on the Killing horizon takes a similar form, although with different component functions $V$, $U$, $U_a$ and $h_{ab}$ to the metric. These must obey the same smoothness conditions, so be smooth functions of $r^2$ and of $x^a$, and also satisfy~\eqref{eq:surfgravcondition}, but need not have the positivity of $V$, and $h_{ab}$ need not be a Riemannian metric.

Now consider a covector $\omega$. The most general form compatible with the symmetry $\mathrm{Lie}_{\partial/\partial t} \omega = 0$, and with smoothness at the Killing horizon can be written as,
\be
\label{eq:smoothcovec}
\omega = r^2 \Phi \, dt + r W dr + \omega_a dx^a
\ee
where now $\Phi$, $W$ and the components $\omega_{a}$ are again smooth functions of $r^2$ and of $x^a$. We may see this by writing a general smooth vector in the coordinates $\alpha, \beta$ as,
\be
\omega = \omega_\alpha d\alpha + \omega_\beta d\beta + \omega_a dx^a
\ee
where $\omega_{\alpha}$, $\omega_{\beta}$, $\omega_a$ are smooth functions of $\alpha$, $\beta$ and $x^a$.
%
%
Then transforming to $t$ and $r$ coordinates we obtain,
\be
\omega = \kappa r \left( \omega_\alpha \cosh{\kappa t} + \omega_\beta  \sinh \kappa t  \right) dt + \left( \omega_\alpha  \sinh \kappa t  + \omega_\beta  \cosh \kappa t \right) dr + \omega_a dx^a \; .
\ee

In order to have the static symmetry we see that these components should have no explicit $t$ dependence. Noting that a smooth function $f$ of $\alpha$ and $\beta$ such that $\partial_t f = 0$ must be a function of $\beta^2 - \alpha^2 = r^2$,
then we see the behaviour of $\omega_\alpha$ and $\omega_\beta$ is constrained to go as,
\be
    \omega_\alpha = - \alpha W(r^2,x^a) + \frac{1}{\kappa}\beta \Phi(r^2,x^a)  \; ,\quad \omega_\beta =  - \frac{1}{\kappa} \alpha \Phi(r^2,x^a) + \beta W(r^2,x^a) 
\ee
for $\Phi(r^2,x^a)$, $W(r^2,x^a)$ smooth functions of $r^2$ and $x^a$, which yields the form in equation~\eqref{eq:smoothcovec} above.

Then a gauge field, $A$, which is purely electric, compatible with the static symmetry and smooth on the Killing horizon will take the form,
\be
A = r^2 \Phi \, dt 
\ee
(after a gauge transformation to remove any spatial components). In the
purely magnetic case, it will take the form,
\be
A =  r W dr + \omega_a dx^a \; .
\ee
We take the reference metric to be some smooth spacetime with a coinciding Killing horizon at $r =0$ with the same surface gravity, so, 
\be
\bar{g}_{\mu\nu} dx^\mu dx^\nu = - r^2 \bar{V} dt^2 + \bar{U} ( dr + r \bar{U}_a dx^a )^2 + \bar{h}_{ab} dx^a dx^b
\ee
so that $\bar{V}$, $\bar{U}$, $\bar{U}_a$ and $\bar{h}_{ab}$ obey the same conditions as the corresponding component functions of the metric. Being static, then staticity with respect to $\partial / \partial t$ is preserved by the flow. Likewise the electric or magnetic form of the gauge potential is also preserved.\footnote{We note that there is no mixing between these electric and magnetic components since they transform differently under the static $t \to -t$ symmetry.}

Translating to $\alpha$ and $\beta$ coordinates, all the tensors are smooth in the coordinates $\alpha$, $\beta$, $x^a$, and compatible with the symmetry $\partial / \partial t$. Now consider the flow equations. The tensors derived from the metric, reference metric and gauge field, the curvature and field strengths and the various Lie derivative and gauge transform terms, will also be smooth and respect the symmetry. Combining these into the `righthand' sides of the flow equations will again yield a smooth symmetric two tensor and a covector. Since these respect the static symmetry, and are smooth in $\alpha$, $\beta$, then returning to $t, r, x^a$ coordinates they will take the smooth forms discussed above, and thus the smooth form of the metric and gauge field will be preserved by the flow, and in particular its constant surface gravity $\kappa$ with respect to $\partial / \partial t$ will be preserved by the flow.

\subsection{Conserved charges under EM flow}
\label{sec:conserved}

Since black holes can carry charges, we may wonder whether any charges or potentials are conserved by the static EM flow. Let us consider firstly the electric flow in asymptotically flat spacetimes. We then take the asymptotic boundary condition that the metric (and reference metric) tend to Minkowski and the non-vanishing component of the vector field, $A^t$, tends to a constant, $\Phi_{\infty}$. Thus we have,
\be
ds^2 \to -dt^2 + dr^2 + r^2 \left( d\theta^2 + \sin^2{\theta} d \phi^2 \right) \; , \quad A^t \to \Phi_{\infty}
\ee
as $r \to \infty$. We may interpret $\Phi_{\infty}$ as the electric  potential difference of the spacetime, since we have seen that $A^t \to 0$ at black hole horizons. Thus we see that for black hole spacetimes the electric static EM flow preserves the electric potential difference, $\mu$, since it is fixed by the asymptotic boundary conditions.

One may ask, does it also preserve the charge?
Working to higher order in powers of $1/r$, given a potential of the form,
\be
A^t = \Phi_\infty + \frac{1}{r} f(\theta,\phi) + O\bigl(\frac{1}{r^2}\bigr)
\ee
then the electric charge is, 
\be
Q = \frac{1}{4 \pi} \int_{S^2_{\infty}} \star F = \frac{1}{4\pi} \int d\theta d\phi \sin\theta f(\theta,\phi)
\ee
where the integral is taken over the 2-sphere at infinite radius.
Naively it appears that this electric charge is also fixed in flow time, since one may simply show that,
\be
\dot{A}^t = O\left( \frac{1}{r^2} \right)
\ee
and thus the leading $1/r$ term in $A_t$ is not affected by the flow. This is similar to the observation in~\cite{Oliynyk:2006nr} that the ADM mass of a static spacetime is invariant under Ricci flow, since again the leading asymptotic fall-off is not corrected.

While technically correct, we argue that this is not physically the correct picture. A counter example in our context that we discussed earlier is that we may Ricci flow from the Schwarzschild black hole to flat spacetime (via a topology change), so clearly we should physically expect that the mass changes.\footnote{One might be tempted to say that this is associated to the surgery required to change topology, but this is not the case as the surgery is local to the region in the interior where the singularity develops as the horizon shrinks, and does not affect the asymptotics at all.}
The resolution is that there are different ways to compute charges or masses during a flow. Charges are defined far away from the system of interest, and thus there are two limits we are concerned with, large radius and large flow times. Depending on the ordering of these limits we may obtain a definition of charge that doesn't vary with flow time, or one that does. If we define charge at fixed finite flow time and infinite radius, we don't allow the charge to vary in a finite flow time as information from the interior of the system cannot propagate out to infinity. On the other hand, if we define charges on a very large but formally finite sphere, these charges may vary at sufficiently late times in the flow. This is analogous to the discussion in~\cite{Oliynyk:2006nr} where although ADM mass is fixed during Ricci flow, quasi-local masses do evolve.

This may be precisely illustrated with a simple example given by the following exact solution to diffusion in flat Euclidean 3d space,
\be
F(\lambda,\vec{x}) = \frac{1}{|\vec{x}|} \int_0^{\infty} d\omega \tilde{F}(\omega) e^{- \omega \lambda} \sin\left(|\vec{x}| \sqrt{\omega}\right)
\ee
where $\vec{x}$ are the usual Euclidean coordinates and $\lambda$ is diffusion flow time, and $\tilde{F}(\omega)$ is the integral transform of the initial data at $\lambda = 0$. We consider an initial profile given by,
\be
\tilde{F}(\omega) = \left\{
\begin{array}{cc}
\frac{1}{\pi \omega} & \omega < 1/L \\
0 & \omega \ge 1/L
\end{array}
\right. 
\ee
and then the $p$-th flow time derivative at $\lambda = 0$ is,
\be
\left. \frac{d^p}{d\lambda^p} F(\lambda,\vec{x}) \right|_{\lambda = 0} = 
\left\{
\begin{array}{cc}
\frac{1}{r} + O\left( \frac{1}{r^2} \right) & p = 0 \\
- \frac{1}{r^2} \frac{2 (-1)^p \cos\left( r/\sqrt{L} \right)}{\pi \sqrt{L}^{2 p -1}}  + O\left( \frac{1}{r^3} \right)  & p > 0 \; .
\end{array}
\right.
\ee
Since $F(0,\vec{x}) \sim 1/r$, but $\left. \frac{d^p}{d\lambda^p} F(\lambda,\vec{x}) \right|_{\lambda = 0} \sim 1/r^2$ for any $p>0$, we see that the asymptotic $\sim 1/r$ behaviour persists unchanged at finite flow time, so that $F(\lambda,\vec{x}) \sim 1/r$.
However it is quite evident that pointwise, so at any fixed position $\vec{x}$, we have $F(\lambda, \vec{x}) \to 0$ as $\lambda \to \infty$.
In diffusion we know that it will take a diffusion time $\lambda \sim r^2$ to effect a change on the scale $r$. Thus while formally at finite flow time the $1/r$ coefficient will remain unchanged, if we actually look at some very large, but finite fixed radius, and ask how the diffusion field looks at late flow times, it will tend to zero, as will all its spatial derivatives. Formally for any point $\vec{x}$,
\be
\lim_{\lambda \to \infty} \partial_{i_1} \partial_{i_2} \ldots \partial_{i_n} F(\lambda,\vec{x}) = 0
\ee
for any number of derivatives $n \ge 0$.

Let us return to the context of the electric static EM flow. The charge is fixed for finite flow times if we compute it from the formula above (ie. from the $1/r$ fall-off). However, if instead we compute it as the charge inside some very large radius $R$,
\be
Q = \frac{1}{4 \pi} \int_{S^2_{r = R}} \star F 
\ee
this will closely approximate the charge computed with the original definition for early times, $\lambda \ll R^2$, but then will change and deviate from this definition if we wait sufficiently long flow times, $\lambda \sim R^2$. With this more physical definition of charge, we conclude that for the electric static EM flow the electric potential is fixed as a boundary condition, and the charge will vary with flow time.

The converse holds for the magnetic static EM flow. Then asymptotically we have the boundary condition that, 
\be
A_a \to A_a(\theta,\phi) 
\ee
for large $r$, where $x^a = \{ \theta, \phi \}$ and $A_a(\theta,\phi)$ is a fixed covector field on the 2-sphere at infinity. We may regard this as a `Dirichlet' boundary condition for the magnetic potential as $r \to \infty$. The magnetic charge is now computed as,
\be
Q_{mag} = \frac{1}{4 \pi} \int_{S^2_{\infty}} F = \frac{1}{4\pi} \int d\theta d\phi \left( \partial_\theta A_\phi - \partial_\phi A_\theta \right)
\ee
and thus is fixed by the asymptotic data $A_a$. As it is the leading behaviour of $A_a$ that determines this magnetic charge in the large $r$ limit, we emphasize that this remains fixed independent of whether we evaluate the charge integral strictly in the infinite radius limit, or instead at some very large but finite radius $R$.

Thus in summary, the electrostatic EM flow preserves the electric potential as a boundary condition (but not electric charge), and the magnetostatic EM flow preserves the magnetic charge as a boundary condition. Both flows preserve the surface gravity. We will return to this point later when drawing an analogy between the behaviour of these flows starting with an RN black hole, and the thermodynamic stability of the RN solution.

\subsection{Static extremal black hole spacetimes}
\label{sec:extremal}

Already for Ricci flow it is far less obvious that the flow is compatible with preserving a static smooth extremal horizon. 

We begin our discussion considering the flow of the metric. Then we will consider adding the gauge field, which is either purely electric or purely magnetic.
Following~\cite{Figueras:2011va} a general static extremal horizon can be written as,
\be
\label{eq:extremal}
ds^2 = - r^2 e^{T} dt^2 + e^{R} \left( \frac{dr}{r} + r \omega_a dx^a \right)^2 + \gamma_{ab} dx^a dx^b
\ee
where $r = 0$ is the horizon and crucially $T$, $R$, $\omega_a$ and $\gamma_{ab}$ are smooth functions of $r$ and $x^a$ there, and further $T$ and $R$ obey,
ì\be
\label{eq:conditions}
(T - R) |_{r=0} = 0 \; , \quad \partial_r (T - R)|_{r=0}  = \psi 
\ee
where $\psi$ is a constant.
The near horizon geometry is then given by the limit as $r \to 0$,
\be
\label{eq:nh}
ds^2_{NH} = - e^T \left( r^2 dt^2 + \frac{dr^2}{r^2} \right) + \gamma_{ab} dx^a dx^b
\ee
which is a (warped) product of AdS$_2$ with the horizon 2-geometry $\gamma_{ab}$. 

Suppose we flow the metric by a symmetric tensor $A_{\mu\nu}$ so that,
\be
\frac{d}{d\lambda} g_{\mu\nu} = A_{\mu\nu} \; .
\ee
Then we should have that the form above is preserved. In order for this to hold the tensor $A_{\mu\nu}$ must have an analogous form to that of the metric,
\be
A_{\mu\nu} = \left(
\begin{array}{ccc}
- r^2  \hat{A}_{tt} & 0 & 0 \\
& \frac{1}{r^2}  \hat{A}_{rr} & \hat{A}_{a} \\
& & \hat{A}_{ab} + r^2  \hat{A}_{a} \hat{A}_{b}
\end{array}
 \right) 
\ee
so that the component functions $ \hat{A}_{tt},  \hat{A}_{rr}, \hat{A}_a, \hat{A}_{ab}$ are smooth in $r$ and $x^a$. Then to preserve the metric smoothness conditions~\eqref{eq:conditions} we further require,
\be
\label{eq:conditions2}
 \left. \left( \hat{A}_{tt} -  \hat{A}_{rr} \right) \right|_{r=0} = 0  \; , \quad \left.  \partial_r  ( \hat{A}_{tt} - \hat{A}_{rr} )    \right|_{r=0} 
 = \psi  \left. \hat{A}_{tt}   \right|_{r=0} \; .
\ee
This tensor is then also smooth on the Lorentzian spacetime. Linearly combining such tensors satisfying~\eqref{eq:conditions2} with the same value of $\psi$ again yields smooth symmetric tensors (also with this value of $\psi$), and contracting their indices with the metric gives a smooth function of $r$ and $x^a$.
The near horizon limit of this smooth tensor exists and is given as,
\be
A^{NH}_{\mu\nu} = \left(
\begin{array}{ccc}
- r^2  A^{NH}(x^a)& 0 & 0 \\
& \frac{1}{r^2}  A^{NH}(x^a) &0 \\
& & A^{NH}_{ab}(x^a)
\end{array}
 \right) 
\ee
where the components are defined from those of the original tensor at $r = 0$ as,
\be
 A^{NH}(x^a) = \left. \hat{A}_{tt}(r,x^a) \right|_{r=0} \; , \quad
 A^{NH}_{ab}(x^a) =  \left. \hat{A}_{ab}(r,x^a) \right|_{r=0} 
\ee
and may be thought of as a scalar and symmetric 2-tensor respectively, defined on the 2-dimensional near horizon spatial geometry $\gamma_{ab}$.

The static EM flow updates both the metric and gauge field by the sum of several terms, and we now consider these individually, showing that smoothness is preserved.

\subsubsection{Ricci tensor term}

We now show that the Ricci tensor of this geometry is smooth. This implies that flowing the metric by the Ricci tensor preserves smoothness. This analysis is simplified by noting that we may use residual coordinate freedom to choose $\omega_a$ to vanish in the metric~\eqref{eq:extremal}. 
Explicitly we may write~\eqref{eq:extremal} as,
\be
ds^2 = - r^2 e^{T} dt^2 + e^{R} \left( 1 - r^2 w_a w^a e^R \right) \frac{dr^2}{r^2} + {\gamma}'_{ab} \left( dx^a + e^R w^a dr \right) \left( dx^b + e^R w^b dr \right)
\ee
where ${\gamma}'_{ab} = \gamma_{ab} + e^R r^2 w_a w_b$ and $w^a$ is defined by $w_a = \gamma'_{ab} w^b$. We now take a new coordinate $y^a(r,x)$ which obeys the linear differential equation,
\be
\label{eq:defny}
\partial_r y^a = e^R w^b \partial_b y^a \; .
\ee
This may be solved by taking boundary conditions $y^a = x^a$ at $r = 0$ and then integrating in $r$ starting from the surface $r = 0$. Since the components of $e^R w^a$ are smooth in $r$ and $x^a$ then this will have smooth solutions and the inverse transformation $y^a \to x^a$ will also exist (at least locally near $r =0$). \footnote{We are grateful to James Lucietti for discussion on this point.}
These coordinates then yield a metric of our desired form,
\be
ds^2 = - r^2 e^{T} dt^2 + e^{R'} \frac{dr^2}{r^2} + \gamma''_{ab} dy^a dy^b \; , \quad \gamma''_{ab} = \gamma'_{cd} \left. \frac{\partial x^c}{\partial y_a} \right|_r \left. \frac{\partial x^d}{\partial y_b} \right|_r \; 
\ee
where $e^{R'}=e^{R}\left( 1 - r^2 w_a w^a e^R \right)$. An important fact we will use later is that since $y^a \simeq x^a$ near $r = 0$, the near horizon metric is unchanged by this transformation. Furthermore the near horizon form of any other smooth tensor is also unchanged.

Now proceeding we will assume we have taken the above coordinates so that our extremal metric can be written as,
\be
ds^2 = - r^2 e^{T} dt^2 + e^{R} \frac{dr^2}{r^2} + \gamma_{ab} dx^a dx^b
\ee
without the off-diagonal $w_a$ terms.
Then by direct calculation one can obtain the Ricci tensor,
\be
R_{\mu\nu} = \left(
\begin{array}{ccc}
- r^2  \hat{R}_{tt} & 0 & 0 \\
& \frac{1}{r^2}  \hat{R}_{rr} & \hat{R}_{a} \\
& & \hat{R}_{ab}  + r^2  \hat{R}_{a} \hat{R}_{b}
\end{array}
 \right) \; .
\ee
We define $K_{ab} = \partial_r \gamma_{ab}$ and $K=\gamma^{ab}K_{ab}$. 
For fixed $r$ then $\gamma_{ab}$ is a 2-metric, whose covariant derivative we denote as $\tilde{\nabla}$, and its Ricci tensor we write as $R^{(\gamma)}_{ab}$. Moreover we  decompose the following expressions covariantly over this $r$-dependent 2-geometry $\gamma_{ab}$ and emphasise that Latin indices are raised/lowered with respect to $\gamma_{ab}$. Then we may write,
\be
 \hat{R}_{tt} &=& e^{T-R} \left( e^R Q_0 + r Q_1 + r^2 Q_2 \right) \nl
 \hat{R}_{rr} &=& \left( e^R W_0 + r Q_1 + r^2 W_2 \right) \nl
\hat{R}_{a} &=& \frac{1}{r} \left( U^0_{a} + r U^1_a \right)  \nl
\hat{R}_{ab}  + r^2  \hat{R}_{a} \hat{R}_{b} & = & e^{-R} \left( e^R V^0_{ab} + r V^1_{ab} + r^2 V^2_{ab} \right)
\ee
where,
\be
Q_0 &=& - \frac{1}{2 } \tilde{\nabla}^2 T - \frac{1}{4 } (\tilde{\nabla} T)^2  - \frac{1}{4 } \tilde{\nabla}_a T \tilde{\nabla}^a R - e^{-R}  \nl
W_0 &=&  - \frac{1}{2 } \tilde{\nabla}^2 R - \frac{1}{4 } (\tilde{\nabla} R)^2  - \frac{1}{4 } \tilde{\nabla}_a T \tilde{\nabla}^a R - e^{-R} \nl
U^0_a &=& - \frac{1}{2} \tilde{\nabla}_a (T - R) 
\ee
and
\be
Q_1 & = & - \frac{3}{2} \partial_r T + \frac{1}{2} \partial_r R - \frac{1}{2} K \nl
U^1_a &=& - \frac{1}{2 } \partial_r \tilde{\nabla}_a T - \frac{1}{4 } \partial_r T \tilde{\nabla}_a (T - R)  + \frac{1}{4 } K \tilde{\nabla}_a R \nl
&& \qquad 
+ \frac{1}{4 } K_{ab} \tilde{\nabla}^b (T - R ) - \frac{1}{2} \tilde{\nabla}_a K + \frac{1}{2} \tilde{\nabla}^b K_{ab}  
\ee
and where,
\be
Q_2 & = & - \frac{1}{2} \partial^2_r T - \frac{1}{4} (\partial_r T)^2 + \frac{1}{4 } (\partial_r T) ( \partial_r R) - \frac{1}{4 } (\partial_r T) K \nl
W_2 & = & - \frac{1}{2 } \partial^2_r T - \frac{1}{4 } (\partial_r T)^2 + \frac{1}{4} (\partial_r T) ( \partial_r R) + \frac{1}{4 } (\partial_r R) K  
- \frac{1}{2 } \partial_r K - \frac{1}{4 } K_{ab} K^{ab} \nl
\ee
and then finally,
\be
V^0_{ab} & = & R^{(\gamma)}_{ab} - \frac{1}{ 2 } \tilde{\nabla}_a \tilde{\nabla}_b T - \frac{1}{ 2 } \tilde{\nabla}_a \tilde{\nabla}_b R - \frac{1}{ 4 } \tilde{\nabla}_a T \tilde{\nabla}_b T - \frac{1}{ 4 } \tilde{\nabla}_a R \tilde{\nabla}_b R \nl
V^1_{ab} & = & - K_{ab} \nl
V^2_{ab} & = & - \frac{1}{2 } \partial_r K_{ab} + \frac{1}{4 } K K_{ab} - \frac{1}{4 }  K_{ab} \partial_r (T - R) + \frac{1}{4 } \gamma_{ab} \left( K^{cd} K_{cd} - K^2 \right) \; . \nl
\ee
Firstly one can check that $\hat{R}_{a}$ is in fact a smooth function of $r$, even though naively it appears to go as $O(1/r)$, due to the fact that $T = R$ at $r = 0$ so $U^0_a$ vanishes at $r = 0$. The other components $\hat{R}_{tt}$, $\hat{R}_{rr}$, $\hat{R}_{ab}$ are then clearly smooth in $r$ and $x^a$. Secondly one can explicitly confirm that the smoothness conditions~\eqref{eq:conditions2} indeed hold, by virtue of the behaviour of the metric functions. 
Furthermore we see that the near horizon geometry of the Ricci tensor is given by,
\be
R^{NH}_{\mu\nu} = \left(
\begin{array}{ccc}
- r^2 R^{NH} & 0 & 0 \\
& \frac{1}{r^2} R^{NH}  &0 \\
& & R^{NH}_{ab}
\end{array}
 \right) 
\ee
where,
\be
R^{NH}  &=& e^{T} \left(  - \frac{1}{2 } \tilde{\nabla}^2 T - \frac{1}{2 } (\tilde{\nabla} T)^2  - e^{-T}  \right) \nl
R^{NH}_{ab} &=& R^{(\gamma)}_{ab} -  \tilde{\nabla}_a \tilde{\nabla}_b T  - \frac{1}{ 2 } \tilde{\nabla}_a T \tilde{\nabla}_b T \; .
\ee
Clearly the near horizon geometry only depends on the intrinsic geometry and not on any $r$ derivatives at the horizon $r = 0$.
Thus we see that the Ricci tensor term in our RN flow equation for the metric indeed preserves the extremal horizon structure.

\subsubsection{DeTurck diffeomorphism term}

Now we consider the diffeomorphism term in the metric flow equation, $\nabla_{(\mu}\xi_{\nu)}$.
In order to keep the metric static and preserve $g_{tr} = g_{ta} = 0$ during the flow, the diffeomorphism vector field must have $\xi^t = 0$, and the remaining components cannot depend on $t$. 
Additionally for the diffeomorphism to preserve the extremal horizon we find it must take the form,
\be
\label{eq:vform}
\xi^\mu = \left( 0, r^2 f , v^a \right)
\ee
where $f$ and $v^a$ are smooth functions of $r$ and $x^a$. The vanishing $\xi^r$ component at $r = 0$ implies that the position of the horizon is not changed in the flow, as one would expect. Perhaps less obvious is that this component must vanish quadratically in $r$ as the following expressions show,
\be
 \nabla_{(\mu} \xi_{\nu)}= \left(
\begin{array}{ccc}
- r^2  \hat{A}_{tt} & 0 & 0 \\
& \frac{1}{r^2}  \hat{A}_{rr} & \hat{A}_{a} \\
& & \hat{A}_{ab}  + r^2  \hat{A}_{a} \hat{A}_{b}
\end{array}
 \right) 
\ee
where we have,
\be
 \hat{A}_{tt} &=& e^{T} \left( \frac{1}{2} v^a \tilde{\nabla}_a T   + r f + \frac{1}{2} r^2 f \partial_r T \right) \nl
 \hat{A}_{rr} &=& e^{R} \left( \frac{1}{2} v^a \tilde{\nabla}_a R   + r f + r^2 \left( \frac{1}{2}  f \partial_r R + \partial_r f \right) \right) \nl
\hat{A}_{a} &=& \frac{1}{2} e^R \tilde{\nabla}_a f + \frac{1}{2} \gamma_{ab} \partial_r v^b  \nl
\hat{A}_{ab}  + r^2  \hat{A}_{a} \hat{A}_{b} & = &  \tilde{\nabla}_{(a} v_{b)} + \frac{r^2}{2} f K_{ab} \; .
\ee
In particular $\hat{A}_a$ would be singular, going as $\sim 1/r$ if $\xi^r \sim r$ rather than quadratically in $r$.\footnote{
We note that while a general behaviour $\xi^r = r f(r, x^a)$ for $f$ smooth in its arguments would be singular, it seems possible to have $\xi^r = r c + r^2 f(r, x^a)$, with $c$ a constant. However as we discuss shortly, the DeTurck vector component $\xi^r$ does not have such a linear in $r$ leading behaviour. Thus, we drop this contribution.
}
One can check that this indeed obeys the smoothness conditions~\eqref{eq:conditions2}. The near horizon form is then,
\be
(\nabla_{(\mu} \xi_{\nu)} )^{NH}= \left(
\begin{array}{ccc}
- r^2  A^{NH} & 0 & 0 \\
& \frac{1}{r^2}  A^{NH} & 0\\
& & A^{NH}_{ab} 
\end{array}
 \right) 
\ee
with,
\be
 A^{NH} = \left. \frac{1}{2}e^{T} v^a \tilde{\nabla}_a T \right|_{r=0} \; , \quad
A^{NH}_{ab} = \left. \tilde{\nabla}_{(a} v_{b)} \right|_{r=0}
\ee
and we see this only depends on $v^a$ and not on $f$.

We may check that the DeTurck choice of diffeomorphism indeed gives the above form for $\xi^\mu$ in~\eqref{eq:vform}. Recall that,
\be
\xi^\mu = g^{\alpha\beta} \left( \Gamma^\mu_{~\alpha\beta} - \bar{\Gamma}^\mu_{~\alpha\beta} \right)
\ee
with the reference connection being the Levi-Civita connection of a smooth static reference metric,
\be
d\bar{s}^2 = - r^2 e^{\bar{T}} dt^2 + e^{\bar{R}} \left( \frac{dr}{r} + r \bar{\omega}_a dx^a \right)^2 + \bar{\gamma}_{ab} dx^a dx^b
\ee
where $\bar{T}$, $\bar{R}$, $\bar{\omega}_a$ and $\bar{\gamma}_{ab}$ obey the same restrictions as their counterparts in the actual metric. In particular, we also need to have $\psi = \bar{\psi}$ in order to obtain a smooth extremal horizon when adding up tensors composed from the actual and the reference metric.
We emphasize that while we have chosen to use our coordinate freedom to eliminate the $dr dx^a$ terms in the metric for these calculations, this generally will not eliminate those terms in the reference metric. 

One then finds the DeTurck vector field indeed has the form required for a diffeomorphism that preserves the extremal horizon, as in equation~\eqref{eq:vform}. It may explicitly be checked that $\xi^t = 0$ and that the component $\xi^r$ goes as,
\be
\xi^r =  r e^{-R} \left( \frac{e^{\bar{T}-\bar{R}}}{ e^{T - R}}  - 1\right) + O(r^2) \; .
\ee
This naively looks to only vanish as $\sim r$ rather than $\sim r^2$ at the horizon. However due to the smoothness of the metric and reference metric, and hence $T - R = \bar{T} - \bar{R} = 0$ at $r = 0$, this expression actually goes as $O(r^2)$, as in equation~\eqref{eq:vform}.
Finally one finds that,
\be
\begin{split}
  \xi^a &= - \frac{1}{2} \gamma^{ab} \tilde{\nabla}_b ( T +  R) 
+ \frac{1}{2} e^{\bar{T}-T} \bar{\gamma}^{ab} \tilde{\nabla}_b \bar{T}\\
&\quad+ \frac{1}{2} e^{\bar{R}-R} \bar{\gamma}^{ab} \tilde{\nabla}_b \bar{R}
+ \gamma^{bc} \left( \tilde{\Gamma}^a_{~bc} - \bar{\tilde{\Gamma}}^a_{~bc}\right)
+ O(r)   
\end{split}
\ee
where $\tilde{\Gamma}^a_{~bc}$ and $\bar{\tilde{\Gamma}}^a_{~bc}$ are the (Levi-Civita) connections of the horizon 2-metric $\gamma_{ab}$, and reference metric $\bar{\gamma}_{ab}$, respectively, and indices are raised/lowered w.r.t. $\gamma_{ab}$.

Hence on the horizon $r = 0$ the 2d vector inducing the near horizon diffeomorphism is,
\be
\label{eq:vNH}
 v^a  =  e^{\bar{T}-T} \bar{\gamma}^{ab} \tilde{\nabla}_b \bar{T}  - \gamma^{ab} \tilde{\nabla}_b T
+ \gamma^{bc} \left( \tilde{\Gamma}^a_{~bc} - \bar{\tilde{\Gamma}}^a_{~bc}\right) \; . 
\ee
Having discussed the Ricci tensor and DeTurck diffeomorphism terms for a static extremal horizon, and shown they give smooth tensors, we now see that DeTurck-Ricci flow will preserve the smooth structure of a static extremal horizon. Moreover we may take the near horizon limit of the flow, to obtain a geometric flow of the near horizon geometry. This near horizon flow will be independent of the details of the geometry away from the horizon, depending only on the intrinsic geometry of the horizon itself.

Now we turn to the remaining parts in the static EM flow. These are the Maxwell field and its flow, and also its backreaction terms in the metric flow equation. We will consider the purely electric and magnetic cases separately.

\subsubsection{The electric Maxwell field}

A purely electric potential is preserved under the EM flow. We find that 
smoothness for the Maxwell field on the static extremal horizon geometry implies that a pure electric potential takes the form, 
\be
\label{eq:Maxwell}
A = r \Phi(r,x^a) dt 
\ee
where the potential function $\Phi$ is smooth in $r$ and $x^a$. Note that the electric potential vanishes at the horizon which is necessary for the vector field $A^\mu$ to have finite norm. Further, as we will see, in order for the field strength $F_{\mu}^{~\alpha} F_{\nu\alpha}$ term in the flow to preserve the 
extremal metric we require,
\be
\label{eq:MaxwellConstraint}
\left. \tilde{\nabla}_a \Phi \right|_{r=0}= 0
\ee
so that $\Phi$ is constant over the horizon.
The right-hand side of the Maxwell field equation in the static EM flow (for a diffeomorphism of the form~\eqref{eq:vform}) can be decomposed as,
\be
\nabla^\mu F_{\mu t } + \mathrm{Lie}_\xi A_t &=& r Q_0 + r^2 e^{-R} Q_1 + r^3 e^{-R} Q_2 \nl
Q_0 & = &  \tilde{\nabla}^2 \Phi - \frac{1}{2} \tilde{\nabla}^a (T - R)  \tilde{\nabla}_a \Phi  + v^a \tilde{\nabla}_a \Phi \nl
Q_1 & = & 2 \partial_r \Phi + \frac{1}{2} \Phi \left( K - \partial_r (T + R) \right) + e^{R} f \Phi \nl
Q_2 & = &  \partial_r^2 \Phi + \frac{1}{2}  \left( K - \partial_r (T + R) \right) \partial_r \Phi + e^{R} f \partial_r \Phi 
\ee
with the $r$- and $a$-components vanishing. 
Note that these terms preserve the form of the electric potential. In particular the condition~\eqref{eq:MaxwellConstraint} is preserved as it implies that $Q_0 = O(r)$. Hence the value of $\Phi$ on the horizon will be constant over the 2-geometry, and remain unchanged in flow time.
Then taking the near horizon limit, the electric gauge potential is just given by,
\be
A^{NH} = r \Phi^{NH} dt
\ee
for constant $\Phi^{NH} = \Phi(0,x^a)$, and is fixed along the flow. It thus generates a constant radially directed electric field, $F^{NH}_{t r}$, on the horizon. The charge enclosed by the horizon is then simply proportional to its horizon area.

We note that the flow gauge transformation term $\partial_\mu \Lambda$ is trivial here because $\Lambda \sim \nabla_\alpha A^{\alpha}$ vanishes for a static electric potential.
Finally the Maxwell field backreacts on the metric flow via the symmetric tensor $F_{\mu~~}^{~\alpha}  F_{\nu\alpha}$ and its trace. We find,
\be
F_{\mu}^{~\alpha} F_{\nu\alpha} = \left(
\begin{array}{ccc}
- r^2  \hat{A}_{tt} & 0 & 0 \\
& \frac{1}{r^2}  \hat{A}_{rr} & \hat{A}_{a} \\
& & \hat{A}_{ab}  + r^2  \hat{A}_{a} \hat{A}_{b}
\end{array}
 \right) 
\ee
where,
\be
 \hat{A}_{tt} &=& -  \tilde{\nabla}_a \Phi \tilde{\nabla}^a \Phi -  e^{-R} \partial_r (r \Phi)  \partial_r (r \Phi)  \nl
 \hat{A}_{rr} &=&  -  e^{-T} \partial_r (r \Phi) \partial_r (r \Phi)   \nl
\hat{A}_{a} &=& - \frac{1}{r} e^{-T} \partial_r (r \Phi)  \tilde{\nabla}_a \Phi  \nl
\hat{A}_{ab}  + r^2  \hat{A}_{a} \hat{A}_{b} & = & -  e^{-T} \tilde{\nabla}_{a} \Phi \tilde{\nabla}_{b} \Phi \; .
\ee
Note we see the previously claimed condition~\eqref{eq:MaxwellConstraint} is required for $( \hat{A}_{tt} -  \hat{A}_{rr})$ to vanish on the horizon, and one can check then that the remaining smoothness constraint in equation~\eqref{eq:conditions2} holds.

Since $F_{\mu}^{~\alpha} F_{\nu\alpha}$ is smooth, its trace with respect to the metric will be too, and so will be the backreaction term $F_{\mu}^{~\alpha} F_{\nu\alpha} - \frac{1}{4} g_{\mu\nu} F^2$ in the metric equation of the EM flow.

\subsubsection{The magnetic Maxwell field}

Now we turn to the purely magnetic case. 
We assume our magnetic Maxwell field to take the following form,
\be
\label{eq:MaxwellMagnetic}
A = b_r(r,x^a) dr+ b_a(r,x^a) dx^a 
\ee
where $b_r$ and $b_a$ are smooth functions in $r$ and $x^a$. This form is preserved by the flow equations, which are given by,
\be
\flows \nabla^\mu F_{\mu r } + \mathrm{Lie}_\xi A_r &=&  Q_0 + r Q_1 + r^2  Q_2 \nl
Q_0 & = & \flows \tilde{\nabla}^a F_{ar} +\frac{\flows}{2} \tilde{\nabla}^a (T-R)F_{ar} +v^a \tilde{\nabla}_a b_r + b_a \partial_r v^a \nl
Q_1 & = & 2 b_r f  \nl
Q_2 & = & f \partial_r b_r + b_r \partial_r f 
\ee
\be
\flows \nabla^\mu F_{\mu a } + \mathrm{Lie}_\xi A_a &=&  Q_0 + r e^{-R} Q_1 + r^2 e^{-R} Q_2 \nl
Q_0 & = & \flows \tilde{\nabla}^b F_{ba} +\frac{\flows}{2} \tilde{\nabla}^b (T+R)F_{ba} + \mathrm{Lie}_{v} b_a
\nl
Q_1 & = & 2 \flows F_{ra}  \nl
Q_2 & = & -\flows\partial_r F_{ra} + \frac{\flows}{2} \partial_r (T-R) F_{ra}+\frac{\flows}{2} K F_{ra}+ \flows\gamma^{bc}K_{ba}F_{cr}\nl
&&+e^R f \partial_r b_a + e^R b_r \tilde{\nabla}_a f
\ee
with the $t$-component being trivial. The gauge fixing term $\partial_\mu \Lambda$ is determined by $\Lambda$ which takes the form,
\be
\Lambda = \flows \nabla_\mu A^\mu &=& \flows\left( \Lambda_0 + r e^{-R} \Lambda_1 + r^2 e^{-R} \Lambda_2 \right) \nl
\Lambda_0 &=& \tilde{\nabla}_a b^a + \frac{1}{2} \tilde{\nabla}^a (T+R) b_a \nl
\Lambda_1 &=& 2 b_r \nl
\Lambda_2 &=& \frac{1}{2} (\partial_r (T-R) +K) b_r + \partial_r b_r 
\ee
so we see this is a smooth function. It is then clear that the covector given by its gradient, $\partial_\mu \Lambda$, then preserves the form of the gauge field~\eqref{eq:MaxwellMagnetic} with its contribution to the flow.
%
%
The term $F_{\mu~~}^{~\alpha}  F_{\nu\alpha}$ gives,
\be
F_{\mu}^{~\alpha} F_{\nu\alpha} = \left(
\begin{array}{ccc}
- r^2  \hat{A}_{tt} & 0 & 0 \\
& \frac{1}{r^2}  \hat{A}_{rr} & \hat{A}_{a} \\
& & \hat{A}_{ab}  + r^2  \hat{A}_{a} \hat{A}_{b}
\end{array}
 \right) 
\ee
where,
\be
 \hat{A}_{tt} &=& 0  \nl
 \hat{A}_{rr} &=&   r^2\gamma^{ab}F_{ra}F_{rb}   \nl
\hat{A}_{a} &=& \gamma^{bc} F_{rb} F_{ac}  \nl
\hat{A}_{ab}  + r^2  \hat{A}_{a} \hat{A}_{b} & = & \gamma^{cd} F_{ac} F_{bd} + r^2 e^{-R} F_{ra} F_{rb}
\ee
with $F_{ra} = \partial_r b_a -\tilde{\nabla}_a b_r$ and $F_{ab} = \tilde{\nabla}_a b_b -\tilde{\nabla}_b b_a$. 
Because we have $\hat{A}_{tt}=0$, and $\hat{A}_{rr} \sim O(r^2)$, the smoothness conditions in \eqref{eq:conditions2} are trivially fulfilled. Hence the backreaction of the magnetic potential in the metric flow preserves smoothness of the extremal horizon.

The near horizon limit of the magnetic gauge potential $A = b_r(r,x^b) dr + b_a(r, x^b) dx^a$ at $r = 0$ becomes the 2-vector potential,
\be
A^{NH} = b^{NH}_a(x^b) dx^a \; , \quad b^{NH}_a(x^b) = b_a(0, x^b)
\ee
defined over the 2-geometry $\gamma_{ab}$ of the horizon.

\subsubsection{Putting the terms together and the near horizon limit}

We have seen above that each term in the static EM flow equations preserves the smoothness of both the extremal horizon and the gauge field. Recall that to simplify the task of computing these terms we have chosen coordinates where the off-diagonal terms, $w_a$, vanish. However we emphasize that having shown smoothness in these simpler coordinates, then guarantees smoothness in more general coordinates where the off-diagonal terms are present. 

We will now consider the near horizon limit of the flow. We explicitly showed above that the transformation to remove these off-diagonal terms does not affect the near horizon form of smooth tensors. Thus having computed the near horizon forms in coordinates where $w_a$ vanishes gives their form for any (smooth) choice of coordinates with terms $w_a$. 

Using the results above we may take the near horizon limit of the EM flow so that for $r = 0$ we have in the electric case,
\be
\frac{d}{d\lambda} T & = & \tilde{\nabla}^2 T + v'^{a} \tilde{\nabla}_a T + 2 e^{-2T} \left( e^T - (\Phi^{NH})^2 \right)\nl
\frac{d}{d\lambda} \gamma_{ab} & = & - 2 R^{(\gamma)}_{ab} + 2 \tilde{\nabla}_{(a} v'_{b)} + \tilde{\nabla}_a T \tilde{\nabla}_b T + 2 \gamma_{ab} e^{-2T} (\Phi^{NH})^2
\ee
where
\be
v'^a = e^{\bar{T}-T}\bar{\gamma}^{ab} \tilde{\nabla}_b \bar{T}  + \xi^{(\gamma) a} 
\; , \quad
\xi^{(\gamma) a} = 
\gamma^{bc} \left( \tilde{\Gamma}^a_{~bc} - \bar{\tilde{\Gamma}}^a_{~bc}\right)
\ee
generates a diffeomorphism along the near horizon flow, and has a contribution from the near horizon 2d DeTurck vector, $\xi^{(\gamma) a}$. We note that, following our discussion above, the near horizon electric potential $\Phi^{NH}$ is simply a constant that doesn't change with flow time. We further note that $v'^a$ differs from the previous $v^a$ above in equation~\eqref{eq:vNH}.

In the magnetic case the near horizon limit of flow takes the following form,
\be
\label{eq:NHmagflow}
\frac{d}{d\lambda} T & = & \tilde{\nabla}^2 T + v'^{a} \tilde{\nabla}_a T + 2 e^{-T}- ( F^{NH} )^2\nl
\frac{d}{d\lambda} \gamma_{ab} & = & - 2 R^{(\gamma)}_{ab} + 2 \tilde{\nabla}_{(a} v'_{b)} + \tilde{\nabla}_a T \tilde{\nabla}_b T + 4 F^{NH}_{ac}F_{b}^{NH~c} - \gamma_{ab} ( F^{NH} )^2\nl
\frac{d}{d\lambda} b^{NH}_a &=& \flows \tilde{\nabla}^b F^{NH}_{ba} + \tilde{\nabla}_a \Lambda^{NH} + \mathrm{Lie}_{v'} b^{NH}_a 
\ee
where $F^{NH}_{ab} = \tilde{\nabla}_a b^{NH}_b - \tilde{\nabla}_b b^{NH}_a$, $\Lambda^{NH} = \flows \tilde{\nabla}^a b_a$, and $v'^a$ is as defined above in the electric case. The two derivative terms on the right-hand side of both the electric and the magnetic metric flow equations have precisely the correct form to give a parabolic flow, 
\be
\frac{d}{d\lambda} T = \gamma^{ab} \partial_a \partial_b T + \ldots \; , \quad 
\frac{d}{d\lambda} \gamma_{ab} = \gamma^{cd} \partial_c \partial_d \gamma_{ab} + \ldots
\ee
where $\ldots$ represent lower derivative terms. Thus $T$ and the metric components obey coupled diffusion equations governed by the near horizon (inverse) metric $\gamma^{ab}$ which is Riemannian, and hence give parabolic flows. We note that it is the 2d near horizon limit of the DeTurck vector that ensures this parabolic flow for the near horizon metric.

For the electric flow the electric potential remains constant. In the magnetic case we see that the principle part of the potential flow is,
\be
\frac{d}{d\lambda} b_{a} = \flows \gamma^{cd} \partial_c \partial_d b_{a} + \ldots
\ee
so that it also yields a parabolic flow, due to the gauge fixing term $\tilde{\nabla}_a \Lambda^{NH}$ in equation~\eqref{eq:NHmagflow}. 

Away from the horizon we know that the EM flow has a parabolic character. It also preserves the smooth static extremal horizon structure. Thus we may construct this near horizon flow and then regard its solution as Dirichlet data for the parabolic flow in the exterior of the horizon.
This gives a heuristic argument for well-posedness in the presence of an extremal horizon. It would be interesting to develop a rigorous mathematical proof of well-posedness.

\section{Spherically symmetric flows about non-extremal Reissner-Nordstr\"om}
\label{sec:RNflows}

Taking the metric to be electric or magnetic Reissner-Nordstr\"om, and taking the reference metric to be the same, gives a fixed point of the EM flow, where $\dot{g}_{\mu\nu} = \dot{A}_\mu = 0$. The first question we now address is to understand the stability of these two fixed points if we initially perturb the metric and Maxwell field from the Reissner-Nordstr\"om form. 
For fixed points that are unstable we further explore where  the instability flows to.
In order to study this we restrict to the spherically symmetric setting. We expect, but emphasize we have no formal argument, that if the spherically symmetric sector of perturbations is stable, then the fixed point will generally be stable.

We write a general smooth static spherically symmetric metric as,
\be
ds^2 = - \rho^2 T(\lambda, \rho) dt^2 + 4 r(\rho)^4 A(\lambda, \rho) d\rho^2  + r(\rho)^2 S(\lambda, \rho)  d\Omega^2 
\ee
where $d\Omega^2 = d\theta^2 + \sin^2{\theta} d\phi^2$, $\rho$ is the compactified radial coordinate introduced in section~\ref{sec:setup} related to the usual coordinate as $r = 1/(1-\rho^2)$. Then $T, A, S$ specify the geometry and  will depend on flow time $\lambda$. Following the discussion in section~\ref{sec:nonextremal}, they should be smooth functions of $\rho^2$ at the horizon $\rho = 0$. 
Flat spacetime in this radial variable $\rho$ takes the form,
\be
ds^2_{flat} = - dt^2 + 4 r(\rho)^4 d\rho^2  + r(\rho)^2 d\Omega^2 \label{eq:flatspace}
\ee
and thus we require $T, A, S \to 1$ as $\rho \to 1$ for our metric above to impose the boundary condition that it is asymptotically flat. Please note that \eqref{eq:flatspace} only covers flat space for $r \ge 1$.

Further we require that,
\be
\left. \left( T - 4 \kappa^2 A \right)\right|_{\rho = 0} = 0 \, .
\ee
Then this spacetime has a smooth non-extremal horizon at $\rho =0$ with surface gravity $\kappa$, and as discussed above, this surface gravity is preserved by the EM flow. Choosing,
\be
T = F \; , \quad A = F^{-1} \; , \quad S = 1 \; ,\quad F(\rho) = 1 - r_- + r_- \rho^2
\ee
then yields the RN metric from section~\ref{sec:setup}. We now make this choice for the reference metric.

Due to static spherical symmetry the most general gauge field must take the form,
\be
A = \rho^2 \Phi(\lambda, \rho) dt + \rho A_\rho(\lambda, \rho) d\rho + A^{(\Omega)} \; , \quad A^{(\Omega)} = A_\theta d\theta + A_\phi d\phi
\ee
where $\Phi$ and $A_\rho$ depend on flow time and are smooth functions of $\rho^2$. Here $A^{(\Omega)}$ is a one-form living on the 2-sphere (so only having $d\theta$ and $d\phi$ components).
In order for the field strength of $A^{(\Omega)}$ to preserve the static spherical symmetry, its components $A_\theta$ and $A_\phi$ must have no time or  radial dependence. Then the only choice compatible with spherical symmetry is the potential, $A^{(\Omega)} = -q_B \cos{\theta} d\phi$  for a constant $q_B$ (or an $SO(3)$ rotation of this). 
As mentioned previously, in spherical symmetry we may consistently have both an electric and magnetic field, as both the electric and magnetic field are radial so there is no Poynting energy flux. However, following our discussion above we will restrict to either consider the purely electric or the magnetic cases. Noting that the radial component is pure gauge, thus we restrict to the forms,
\be
A_{elec} = \rho^2 \Phi(\lambda,\rho) dt \; , \quad  A_{mag} = - q_B \cos{\theta} d\phi  \; .
\ee
with $q_B$ a constant. We observed previously that in the magnetic case, magnetic charge is fixed along the flow by boundary conditions, and here we see it is given by $q_B$. In the electric case we require $\Phi(\rho) \to \mu$ as $\rho \to 1$, and then $(G \mu)$ gives the electric potential difference between the horizon and infinity (since the potential $A_t \sim \rho^2$ vanishes at the horizon).

We see there is a fundamental difference between the electrostatic and magnetostatic EM flows, namely that in the magnetic case there is no dynamics associated to the gauge field, whereas in the electric case the potential $\Phi(\rho)$ flows. 
Is this fixed form  in the magnetic case consistent with the EM flow equations? This magnetic potential obeys $\nabla^\mu F_{\mu\nu} = 0$ for any choice of metric functions $T$, $A$ and $S$. The additional terms in the Maxwell part of the EM flow, $\partial_\mu \Lambda$ and $\Lie_\xi A_\mu$, both vanish in this magnetic case, since $\Lambda = \flows \nabla_\alpha A^\alpha$  vanishes, and $\xi$ is radially directed (taking the RN reference metric above) but $A_\mu$ has no radial dependence. Thus indeed the magnetic form is held invariant under the flow.

In the electric case we may solve $\nabla^\mu F_{\mu\nu} = 0$ on the static spherically symmetric metric above by taking,
\be
\label{eq:elecpotl}
 \Phi(\rho) = - \frac{q}{\rho^2} \left( \int_0^\rho d\rho' \, \rho' \, \frac{\sqrt{T(\rho') A(\rho')}}{S(\rho') } \right) 
\ee
which then has charge $q$. However, clearly if we deform the geometry, changing $T$, $A$, $S$ as will happen as the metric flows, then this will no longer yield vanishing $\nabla^\mu F_{\mu\nu}$, and hence the electric potential will flow.
Thus we see that the dynamics of electric and magnetic gauge potentials are quite different under the EM flow.

\subsection{Linear perturbations about RN}

We now consider linear perturbations to both these fixed points.
To proceed we expand around the fixed point RN geometry and look for EM flow solutions of the form,
\be
g_{\mu\nu} = g^{(RN)}_{\mu\nu} + \epsilon e^{\Omega \lambda} \delta g_{\mu\nu} \; , \quad A_{\mu} = A^{(RN)}_{\mu} +  \epsilon e^{\Omega \lambda} \delta A_{\mu}
\ee
linearizing in $\epsilon$, where $\Omega$ is a complex constant, and $\delta g_{\mu\nu}$ and $\delta A_\mu$ preserve the static spherical symmetry. 
We expect all such linear `mode' solutions with this exponential flow time dependence provide a basis for the general flow solution, and we may regard finding these solutions as an eigenvalue problem, where $\Omega$ is the eigenvalue. 
Mode solutions with $\mathrm{Re}(\Omega) \le 0$ are stable in flow time, and unstable modes have  $\mathrm{Re}(\Omega) > 0$. 
The existence of such unstable modes would indicate the RN fixed point is an unstable fixed point of the static EM flow, in the sense that a generic initial perturbation to the RN fixed point will contain some component of the unstable mode, and this will take the flow away from the fixed point as it exponentially grows in flow time.

We will choose the reference metric to be the same RN solution as that of the fixed point metric, and hence $\xi^\mu = 0$ at the fixed point. In both the electric and magnetic cases we also have $\Lambda =0 $ for the RN solution with the gauge potentials discussed previously. Thus perturbing the RN solution as above, we will have,
\be
\xi^\mu = \epsilon e^{\Omega \lambda}  \delta \xi^\mu \; , \quad \Lambda = \epsilon  e^{\Omega \lambda}  \delta \Lambda \; .
\ee
We may then separate such eigenmodes into those with $\delta \xi^\mu = \Lambda = 0$, and those with $\delta \xi^\mu, \Lambda \ne 0$. The former case will be an eigenmode on the original Einstein-Maxwell flow equation~\eqref{eq:EMflowNoDT}, whilst the latter is an eigenmode up to a flow time dependent diffeomorphism and gauge transformation.

More explicitly we take,
\be
T(\rho) &=& \left( 1 - r_- + r_- \rho^2  \right) \left( 1 + \epsilon e^{\Omega \lambda} \delta T(\rho) \right) \\
A(\rho) &=& \left( 1 - r_- + r_- \rho^2  \right)^{-1} \left( 1 + \epsilon e^{\Omega \lambda} \delta A(\rho) \right) \\
S(\rho) &=& 1 + \epsilon e^{\Omega \lambda} \delta S(\rho) 
\ee
and in the electric case, we take the electric potential as,
\be
\Phi(\rho) &=& \sqrt{r_-} + \epsilon e^{\Omega \lambda} \delta \Phi(\rho)  
\ee
while in the magnetic case, as discussed above, the gauge potential is simply fixed, with magnetic charge $q_B = \sqrt{r_-}$.

We now proceed to study the unstable modes of the flow equations numerically over the interval $\rho \in [0,1)$.
The functions, $\delta T, \delta A, \delta S$, and in the electric case $\delta \Phi$ too, are all smooth functions of $\rho^2$ (so even functions) on $\rho \in [0,1)$. We then discretize the system of equations in terms of the function $\delta T, \delta A, \delta S$ and $\delta \Phi$ using even pseudospectral differencing~\cite{Trefethen} which automatically imposes the correct even behaviour at the horizon $\rho = 0$. We further impose the asymptotically flat boundary condition that $\delta T = \delta A = \delta S = \delta \Phi =0$ at $\rho = 1$. 
Having discretised, our flow equation is then schematically of the form,
\be
O v = \Omega v
\ee
where due to the discretization $O$ and $v$ are a matrix and vector respectively, and $v$ represents the values of the functions $\delta T, \delta A, \delta S$, and in the electric case $\delta \Phi$, at the discrete points in the interior of the interval $[0, 1]$. 
If we take an even discretization of $[0, 1]$ with $N$ points, not including the boundary at $\rho = 1$, then $O$ is an $N \times N$ matrix, and $v$ is an $N$-vector. We find the mode solutions  by numerically determining the eigenvalues (ie. the values of $\Omega$) and eigenfunctions of the matrix $O$. At any fixed resolution only the lower lying eigenvalues represent eigenmodes of the continuum flow, with large eigenvalue modes being discretization artefacts. Since the negative modes we will find lie at modest values of $| \Omega |$, it is straightforward to find them with relatively low resolution. We may check from the wavefunctions, and by increasing resolution that they are indeed continuum modes, and also assess the numerical error in the eigenvalues we determine. We find that already 100 points gives excellent accuracy for the eigenvalues of these negative modes (to better than one part in $10^{9}$).

\subsection{Magnetic non-extremal Reissner-Nordstr\"om linear stability}

We now consider the eigenfunctions of the flow linearized about magnetic RN. Since the gauge field has no dynamics, the constant $\flows$ plays no role.
The work of Monteiro and Santos~\cite{Monteiro:2008wr} shows that Euclidean magnetic RN has a negative mode for $| Q | < \sqrt{3}G M/2$, which for us implies $r_- < 1/3$. This mode involves the metric, and not the gauge potential which is simply the unperturbed magnetic solution $A_{mag} = - \sqrt{r_-}  \cos{\theta} d\phi$. Continuing back to Lorentzian signature, these negative modes precisely yield $\delta \xi = 0$ unstable eigenmodes of the linearized static EM flow equations for the metric,  with real $\Omega > 0$, signifying an exponentially growing instability.  The fact that $\delta \xi = 0$ for these modes originates from the fact that the tensor and vector perturbation decouple in this magnetic case~\cite{Monteiro:2008wr}.
One might naively then think that the magnetic RN solution becomes a stable fixed point for larger charges and up to extremality. 
However, it isn't clear that unstable modes cannot exist outside the sector $\delta \xi = 0$, as this has not previously been studied. We emphasize that while such modes would not have a clear interpretation in terms of the Euclidean action studied in~\cite{Monteiro:2008wr}, if present they would certainly affect the stability of the fixed point of the static EM flow and that is our concern here.

\begin{figure}
\centerline{
  \includegraphics[width=8cm]{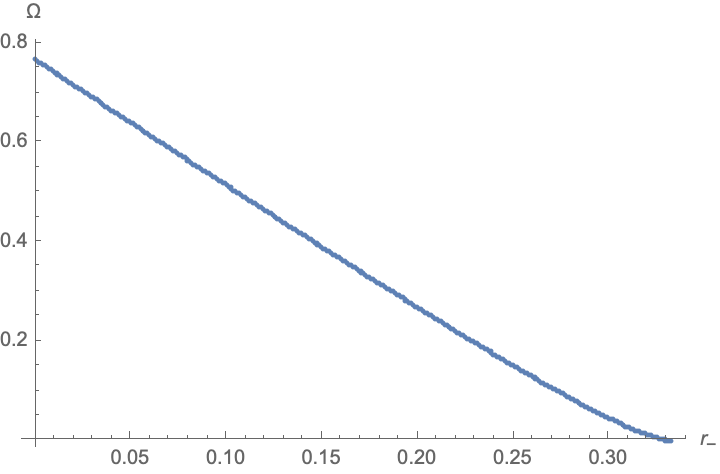}  \includegraphics[width=8cm]{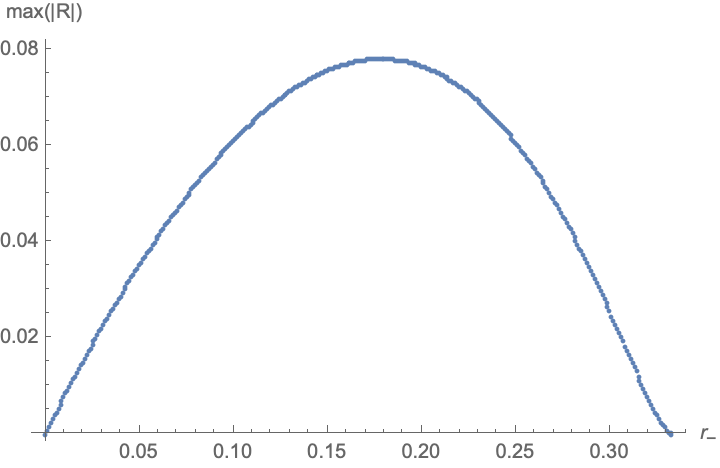}
  }
  \caption{\label{fig:MagNeg}
  The left panel shows the eigenvalue, $\Omega$, of the one unstable mode for the flow of the non-extremal magnetically charged RN solution against $r_-$ which controls its charge. For $r_- = 0$ this is the Gross-Perry-Yaffe negative mode of Schwarzschild. For charges $0 < r_- < 1/3$ this unstable mode persists, disappearing at $r_- = 1/3$, when it becomes a zero mode which is generated by a perturbation within the RN family of solutions. For greater charges the fixed point is stable (within this spherically symmetric setting). $\xi$ remains zero for all these unstable modes, and in the right panel the maximum value of the perturbation to the Ricci scalar for the mode is shown. 
    }
\end{figure}

\begin{figure}
\centerline{
  \includegraphics[width=8cm]{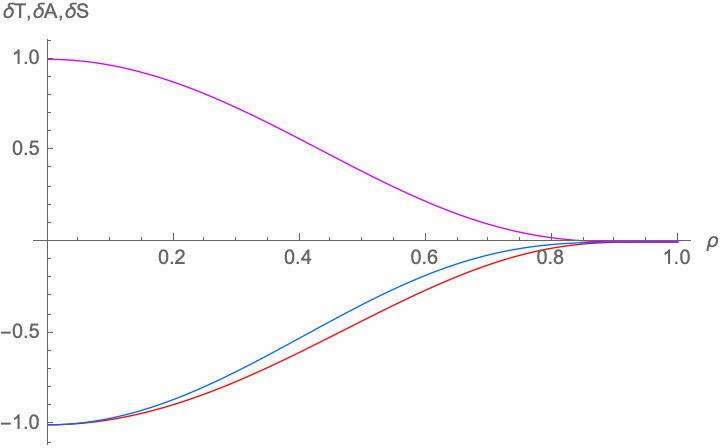}  \includegraphics[width=8cm]{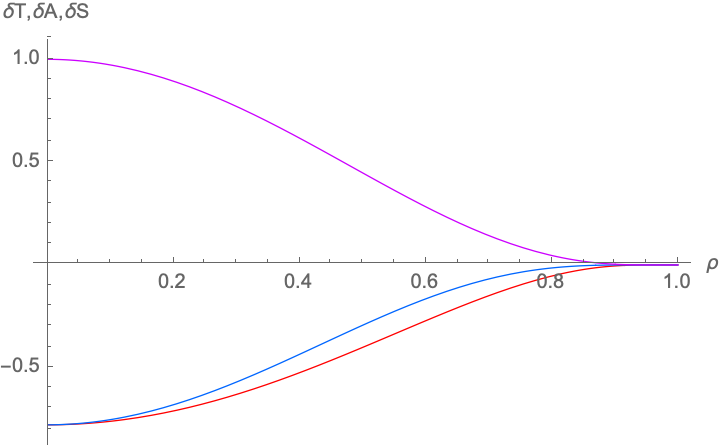} \\
  }
  \centerline{
  \includegraphics[width=8cm]{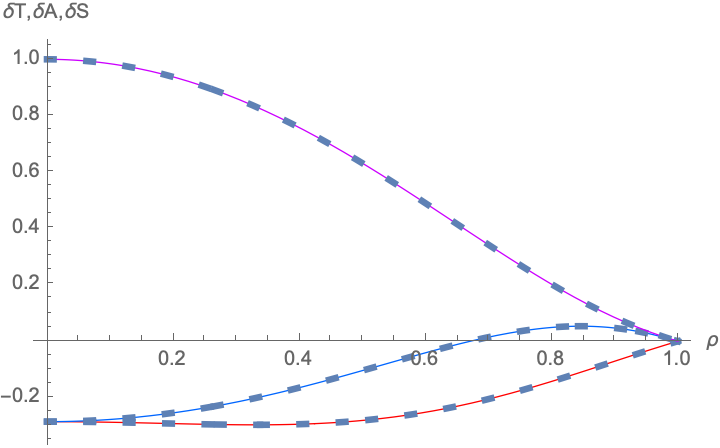} 
  }
  \caption{\label{fig:MagWavefn}
  The numerically generated wavefunctions, $\delta T$ (Red), $\delta A$ (Blue) and $\delta S$ (Purple) for the unstable mode are plotted against the radial coordinate $\rho$ for the unstable mode for $r_- = 0$ (top left),  $r_- = 1/6$ (top right) and $r_- = 1/3$  (bottom). The instability becomes a static perturbation for $r_- = 1/3$, which corresponds to a perturbation within the RN family, and may be computed independently. This is compared by means of dashed curves in the bottom plot, the  agreement providing a check of our numerically constructed eigenfunction.
    }
\end{figure}

In figure~\ref{fig:MagNeg} we show the unstable mode determined numerically against varying charge. We find that there is only one unstable mode for $r_- < 1/3$, and we find none for $r_- > 1/3$. We explicitly check the numerical wavefunctions of the unstable mode for charges where it exists and see that it is compatible with $\delta \xi = 0$ in the continuum. Also plotted in that figure is the maximum value of the perturbation to the Ricci scalar. For the unstable mode of Schwarzschild, so for $r_- = 0$ this vanishes as the Euclidean negative mode is traceless. We see it is non-vanishing for finite charge, but vanishes again when the unstable mode disappears at the critical $r_- = 1/3$, since as we will argue shortly, at that point the mode is a static perturbation of the Einstein-Maxwell equations and so must have vanishing Ricci scalar.
 In figure~\ref{fig:MagWavefn} we plot the wavefunctions $\delta T$, $\delta A$, $\delta S$ for $r_- = 0$, $r_- = 1/6$ and $r_- = 1/3$, normalized so that $\delta S = 1$ at the horizon. This data is for $N = 400$ points, and we find $| \xi | < 2 \times 10^{-6}$ on the interval in all these cases.

Thus we conclude that the only unstable mode in the static spherically symmetric sector is the continuation of the Euclidean mode described by Monteiro and Santos, and having  $\delta \xi = 0$. It is possible that one exists without spherical symmetry, but we believe it would be unlikely.

Finally we address what the perturbation mode is at the critical charge, $r_- = 1/3$. This was not discussed explicitly in~\cite{Monteiro:2008wr}. Since it has $\Omega = 0$ then it is a static perturbation of the RN solution. Given the uniqueness of RN this static perturbation must be generated by a deformation in the family of RN solutions. Indeed we can confirm that this is the case. For $r_- = 1/3$ we can perturb the RN parameters $r_\pm$ and also perform a coordinate transformation as follows,
\be
r_+ = 1 + \epsilon \delta r_+ \; , \quad r_- = \frac{1}{3} + \epsilon \delta r_- \; , \quad \rho \to \rho \left( 1 + \epsilon f(\rho) \right) \; .
\ee
Further we wish to preserve the charge, $Q = \sqrt{r_+ r_-}$ which implies,
\be
\delta r_- = - \frac{1}{3} \delta r_+ \; .
\ee
This deformation within the RN magnetic solution family, together with the coordinate transformation, then induces a perturbation to our fixed point,
\be
\begin{split}
     T &=  \left( 1 -  \frac{1}{3} (1 - \rho^2) \right) \left( 1 + \epsilon \delta T \right) \; , \quad  A =  \left( 1 -  \frac{1}{3} (1 - \rho^2) \right)^{-1} \left( 1 + \epsilon \delta A \right) \; ,\\  S &= 1 + \epsilon \delta S \nl
\end{split}
\ee
with,
\be
\delta T = \frac{\frac{2}{3} (1 - \rho^2 )  \delta r_+ + 2 \left( 1 - \frac{1}{3} ( 1 - 2 \rho^2 ) \right) f(\rho) }{1 - \frac{1}{3} (1 - \rho^2)} \; , \quad \delta S = 2 \delta r_+ + \frac{4 \rho^2 f(\rho)}{1 - \rho^2} 
\ee
and a similar expression for $\delta A$. 
We take $\delta T$, $\delta A$, $\delta S$ from our numerically constructed zero mode, which recall we have normalized so that $\delta S = 1$ at the horizon.
Since for our perturbation above, $\delta S = 2 \delta r_+$ at $\rho = 0$, thus this perturbation is generated by a deformation of the RN with $\delta r_+ = 1/2$. The coordinate transformation is determined by the condition that $\xi = 0$, which implies that $f$ obeys,
\be
f'' + \frac{3 + 5 \rho^2}{\rho ( 1 - \rho^2)} f' + \frac{8( 3 + 4 \rho^2 - \rho^4)}{2 - 3 \rho^2 + \rho^6} f = - \frac{4 \delta r_+}{2 +  \rho^2} \; .
\ee
We require the behaviour that $f \sim f_0 + O(\rho^2)$, so $f$ gives a smooth metric perturbation at $\rho = 0$ (the other behaviour $f \sim 1/\rho^2$ gives a non-smooth metric perturbation). One finds $f \sim (1-\rho)$ as $\rho \to 1$, and we require $f'(1) = \delta r_+$ in order to have $\delta A = \delta S = 0$ at $\rho = 1$ which is imposed by our boundary conditions. One may solve this o.d.e. with these boundary conditions numerically, and  taking $\delta r_+ = 1/2$ to match to our numerically determined zero mode, we obtain $f_0 \simeq - 0.393$, and a numerical form of $f(\rho)$ on $\rho \in [ 0, 1)$. From this we can then deduce the form of $\delta T$, $\delta A$ and $\delta S$ from the above expressions, and we may check these match those of the numerically determined $\Omega = 0$ mode. Indeed we find they match precisely up to numerical error. These wavefunctions, determined from $f$ above are shown in figure~\ref{fig:MagWavefn} as the dashed curves, and those of the numerical static mode are shown as thin solid lines. We see they lie on top of each other, confirming this mode is indeed a perturbation within the RN family.

\subsection{Electric non-extremal Reissner-Nordstr\"om  linear stability}

Now we consider the fixed point given by the electrically charged RN solution. As discussed above, there is no decoupling of the metric flow from that of the gauge field electric potential. In particular one cannot perturb the metric and consistently have the gauge field fixed -- it will be forced to flow. 
Furthermore since the linearized flow will depend explicitly on the parameter $\flows$, this may influence the presence of unstable modes. For simplicity we will focus on the case of $\flows = 1$ here.
In this electric case there is no straightforward Euclidean continuation, and therefore no prior results from considering Euclidean negative modes. 

\begin{figure}
\centerline{
  \includegraphics[width=10cm]{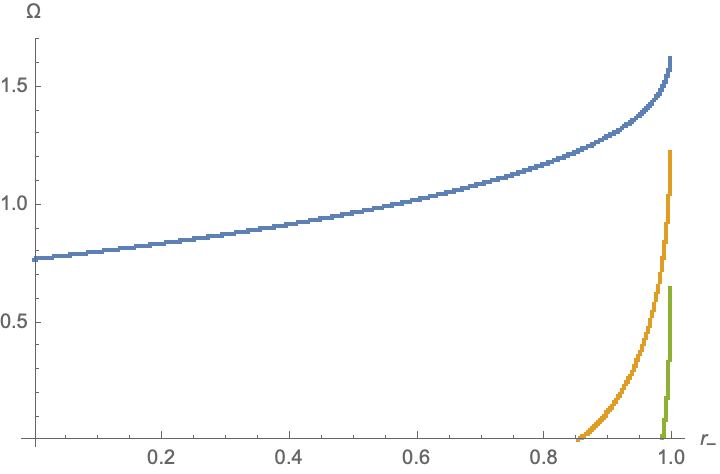} 
  }
  \caption{\label{fig:Elec}
  Figure showing the eigenvalues, $\Omega$, for the unstable modes for the flow of the electrically charged RN solution against $r_-$ for $r_- \le 0.999$ (and for Einstein-Maxwell flow parameter $\flows = 1$). The GPY negative mode at $r_- = 0$ continues to an unstable mode of the electrically charged solutions. However additional negative modes appear as the charge is increased. For $r_- \le 0.999$  up to three unstable modes may exist, but very near extremality, for $0.999 < r_- < 1$ (not shown in the plot) more unstable modes  appear. It therefore appears that electrically charged RN is an unstable fixed point of the EM flow near extremality, unlike the magnetically charged RN solution which is stable.
    }
\end{figure}

For $\flows = 1$ it appears that at least one unstable mode is present up until extremality. Perhaps more surprisingly new unstable modes appear at specific values of the electric charge. 
At the numerical resolutions we use it is difficult to see the behaviour very near extremality, but it appears that an increasing number of unstable modes appear in the limit that $r_- \to 1$. 
In figure~\ref{fig:Elec} we plot the eigenvalues with positive real part, noting that in fact we find all these eigenvalues are purely real. Their wavefunctions are real and we have also confirmed that they are compatible with a smooth continuum form upon increasing resolution. All the unstable modes here, except in the zero charge $r_- = 0$ case, have non-vanishing $\delta \xi$, and non-trivial gauge potential $\delta \Phi$. We have checked that they all have a non-vanishing perturbation to the Ricci scalar, confirming that they are not simply a pure diffeomorphism on the metric, but physically change the geometry as well as the gauge field.

\begin{figure}
\centerline{
  \includegraphics[width=8cm]{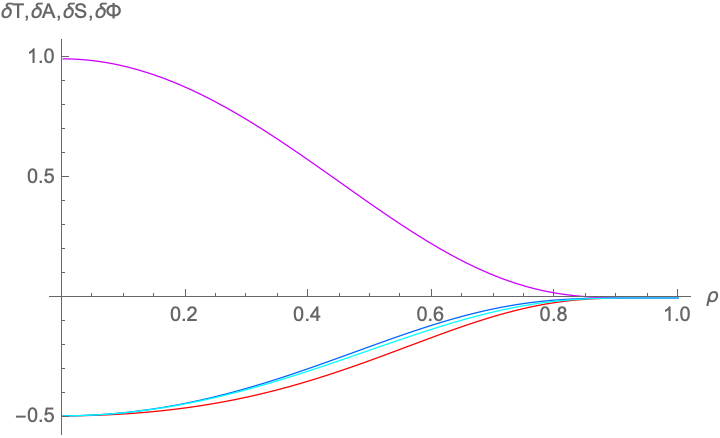}  \includegraphics[width=8cm]{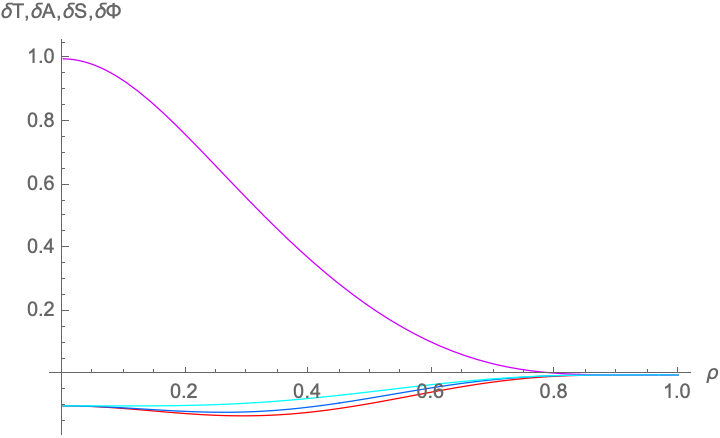} \\
  }
  \centerline{
  \includegraphics[width=8cm]{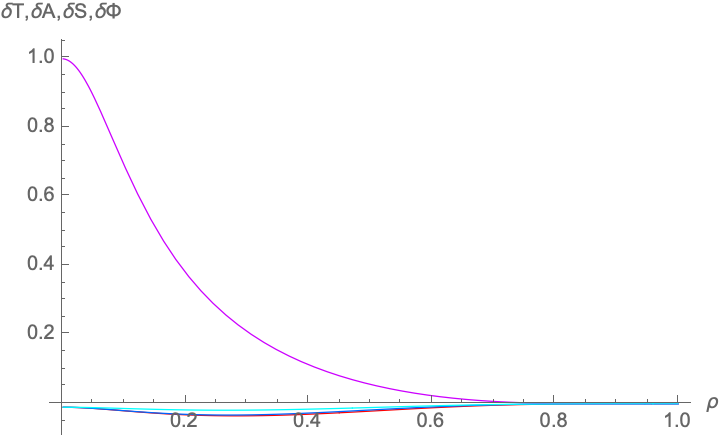}   \includegraphics[width=8cm]{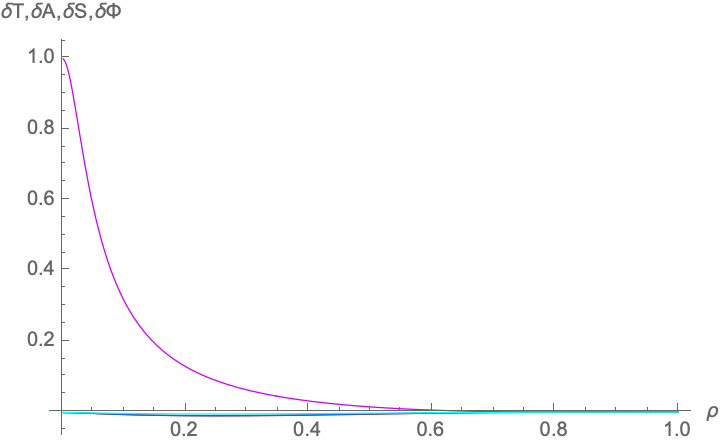} 
  }
  \caption{\label{fig:ElecWavefn1}
 Figure showing the most unstable mode wavefunctions for flow of perturbations of electrically charged RN for charges $r_- = 1/2$ (top left), $r_- = 0.9$ (top right), $r_- = 0.99$ (bottom left) and $r_- = 0.999$ (bottom right). The wavefunction $\delta T$ is shown in red, $\delta A$ in blue, $\delta S$ in purple, and the potential perturbation $\delta \Phi$ is in light blue. The modes are normalized to have $\delta S = 1$ at the horizon, and we see $\delta S$ dominates the near horizon behaviour as extremality is approached, and becomes increasingly localized there.
 }
\end{figure}

\begin{figure}
\centerline{
  \includegraphics[width=8cm]{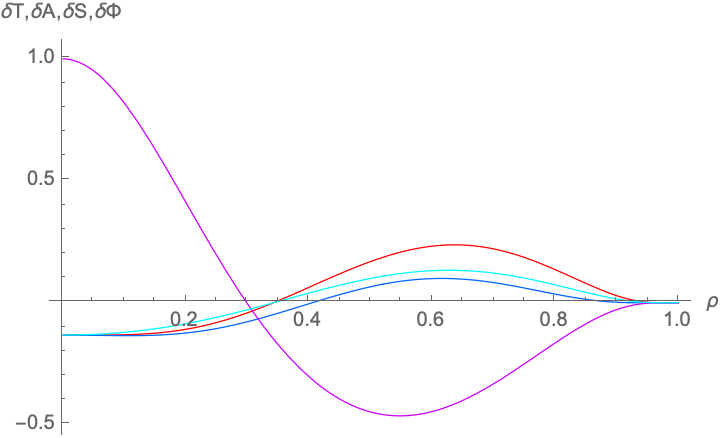}  \includegraphics[width=8cm]{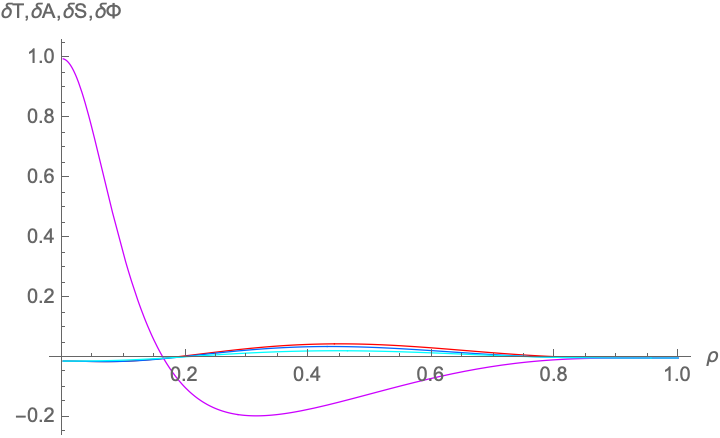} \\
  }
  \centerline{
   \includegraphics[width=8cm]{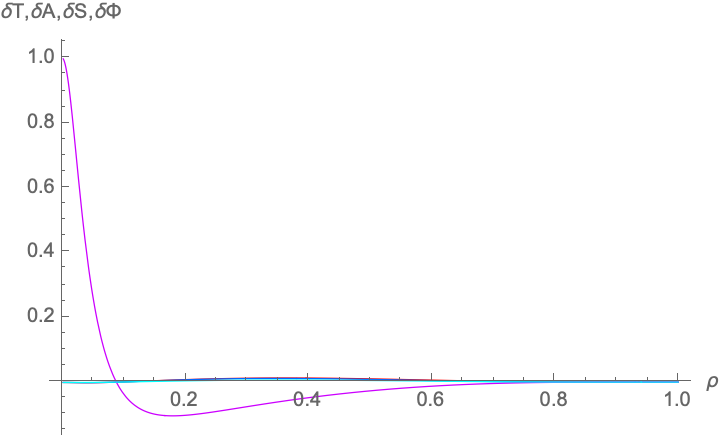} \\
  }
  \caption{\label{fig:ElecWavefn2}
Similar to the previous figure, the wavefunctions for the second most unstable mode are shown for perturbations of electrically charged RN, for $r_- = 0.9$ (top left), $r_- = 0.99$ (top right) and $r_- = 0.999$ (bottom). We see the wavefunctions have one node, and again $\delta S$ grows relative to the other component functions at the horizon as extremality is approached. 
 }
\end{figure}

In figure~\ref{fig:ElecWavefn1} and~\ref{fig:ElecWavefn2} we show the wavefunctions for values $r_- = 0.5, 0.9, 0.99, 0.999$ of the two most unstable modes present for those values (the second unstable mode only emerges for $r_- > 0.854$). We note that the number of nodes increases with each new branch of unstable modes. In the limit $r_- \to 1$ the wavefunctions appear to tend only to have non-trivial $\delta S$ with the other components and gauge field apparently vanishing in this limit. We also observe that the curvature of the function $\delta S$ at the horizon appears to increase as $r_- \to 0$, with $\delta S$ becoming increasingly localized at the horizon $\rho = 0$, indicating the perturbation may not be smooth in the limit of extremality.

Finally we comment that the emergence of new unstable modes with increasing charge indicates that at special values of charge, there are static modes, so linear perturbations that don't flow. This occurs in the magnetic case when the unstable mode disappears, but there $\delta \xi =0$ and at the critical charge this zero mode is just a perturbation tangent to the RN space of solutions as discussed above.
Here however $\delta \xi \ne 0$, and we believe that these zero modes are not associated to perturbations that are tangent to the RN solution. Instead they should be tangent to new branches of `soliton' solutions of the Einstein-Maxwell system that presumably merge with the RN solutions at particular values of charge (depending on the parameter $\flows$). 
We have not attempted to directly construct these soliton solutions  as non-linear solutions to the Einstein-Maxwell soliton equation, but it should be possible to do so.

\subsection{End point of unstable magnetic and electric flows}

In the case of the uncharged asymptotically flat Schwarzschild solution in static spherical symmetry, the single unstable mode of Ricci flow generates two flows, one that leads the horizon to shrink to zero size in finite flow time, and the other that expands the horizon `eating up' the whole spacetime~\cite{Headrick:2006ti}. In both the magnetic case for subcritical charge and in the electric case for all charges we have instabilities and we now ask where these flow.

In order to address this we solve the non-linear flow described at the start of this section~\ref{sec:RNflows}. We evolve the metric functions $T, A, S$ in the ansatz above, and for the electric case also the potential $\Phi$. We discretize the $\rho$ coordinate on the interval $[0,1]$ using sixth order finite differencing. We impose a Neumann boundary condition for the functions $T, A, S$ and $\Phi$ at $\rho = 0$ in order to enforce the smooth horizon boundary condition (which requires these functions to be smooth in $\rho^2$). The asymptotic flatness is imposed as a Dirichlet boundary condition, $T, A, S \to 1$ at $\rho = 1$. In the electric case the potential $\Phi \to \sqrt{r_-}$ at $\rho = 1$, fixing the potential difference to the horizon. We evolve the parabolic system using second order accurate Crank-Nicholson differencing in flow time. The resulting implicit system is solved using an iterative method. We have checked that the code correctly converges to the continuum. The results presented here are for $N = 200$ spatial points, and a flow time step of $1 \times 10^{-5}$ or less. We have also checked that the horizon smoothness and the value of the surface gravity are correctly preserved by the numerical flows.
We reiterate that we have only considered the flow parameter $\flows = 1$ here. 

We discuss the electrostatic EM flow first. We begin by perturbing the RN fixed point, taking initial data,
\be
T = 1 - (1 - \rho^2) r_- \; , \quad A = 1/T \; , \quad S = 1 \pm 0.01 \times (1 - \rho^2)^4 \; , \quad \Phi = \sqrt{r_-}
\ee
so we are perturbing only the sphere part of the metric via the function $S$. 
Whilst obviously finite, this deformation to $S$ is sufficiently small in amplitude that its initial evolution is well described by linear perturbation theory about the RN solution. It preserves the boundary conditions, and depending on the sign either initially decreases ($-$ sign) or increases  ($+$ sign) the horizon size. We expect it to have overlap with the linear unstable mode (or modes) of the electric RN fixed point, and thus to generate flows driven by these unstable modes.

\begin{figure}
\centerline{
  \includegraphics[width=8cm]{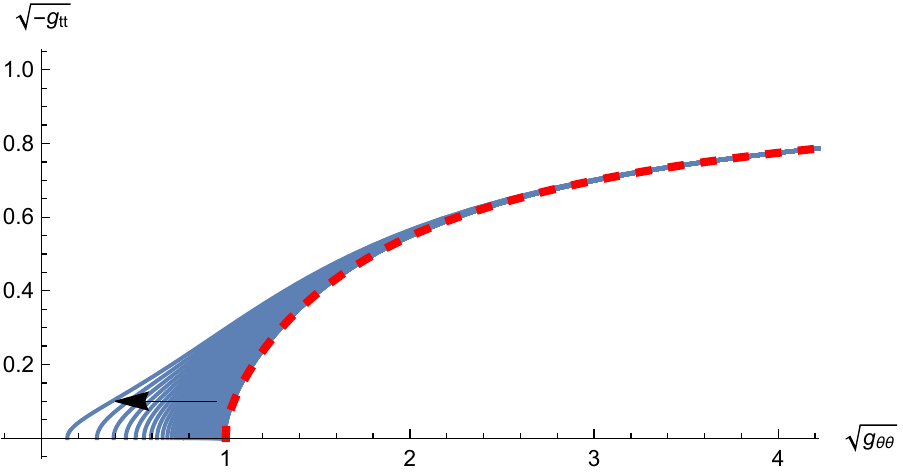}   \includegraphics[width=8cm]{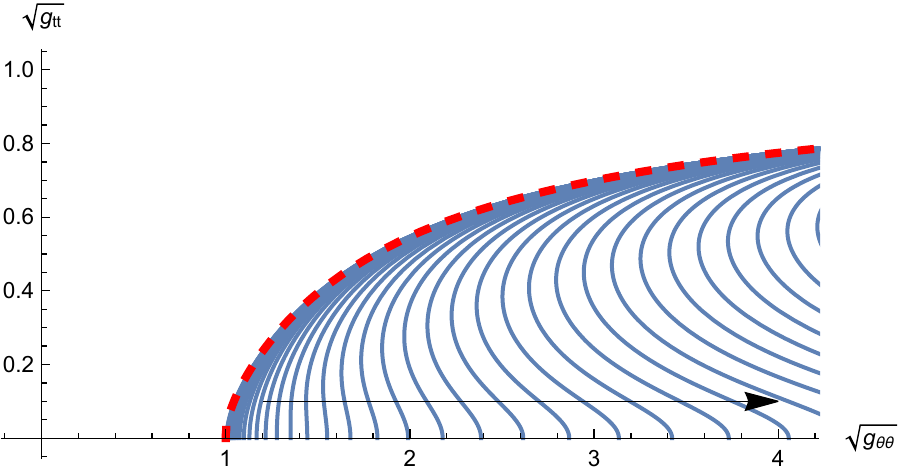} 
}
  \caption{\label{fig:NonExtElec}
  Figure showing curves of $\sqrt{-g_{tt}}$ plotted against $\sqrt{g_{\theta\theta}}$ at constant intervals of electric EM flow time $\Delta \lambda$ and for a small initial perturbation of an electric RN solution with reasonably large charge corresponding to $r_- = 0.8$ (recall $r_+ = 1$). The lefthand frame shows the flow where the horizon is initially reduced in size, and we see that it flows to a singularity at $\lambda \simeq 3.9$ (here $\Delta \lambda = 0.05$). The righthand frame shows the situation where the horizon is perturbed initially to be larger than that of RN, and we see it apparently expands forever (here the curves are shown for flow intervals $\Delta \lambda = 0.25$). 
  The geometry of the RN fixed point is shown by the red dashed curve, and the arrows indicate the sense of change of the geometry for increasing flow time.
  Both these behaviours are qualitatively similar to the Schwarzschild fixed point under Ricci flow. We see analogous behaviour for all charges of the electric RN fixed point.
    }
\end{figure}

In figure~\ref{fig:NonExtElec} we show the evolution of the geometry by plotting the invariants $\sqrt{-g_{tt}} = \rho \sqrt{T}$ against  $\sqrt{g_{\theta\theta}} = r(\rho) \sqrt{S}$ at a sequence of flow times $\lambda$. These are invariants in the sense that they transform as scalars under a gauge transformation $\rho \to F(\rho)$. The curve of $\sqrt{-g_{tt}}$ against $\sqrt{g_{\theta\theta}}$ is then gauge invariant and illustrates how $g_{tt}$ varies with areal radius.
In the figure we show the two evolutions started with the negative and positive initial perturbations to the horizon radius for a reasonably large charge, corresponding to $r_- = 0.8$. We see the perturbation that initially shrinks the horizon flows in finite flow time to a singularity (here occuring at $\lambda \simeq 3.9$), whereas the perturbation that initially expands the horizon leads to it growing in flow time in an apparently unbounded manner. 
In the latter case the horizon accelerates in its expansion, and we can only follow it numerically for a finite period, before time gradients become too large to resolve accurately. 

This behaviour is qualitatively the same behaviour as for the uncharged Schwarzschild solution. We see precisely the same qualtitative behaviour for smaller and larger charges, although as one approaches near extremal charges it becomes more challenging to perform the simulations. For large charges where linear theory tells us multiple unstable modes exist, presumably one may seed the initial flows by these different unstable modes to generate different flows. However we expect that without fine tuning it is the dominant unstable mode that determines the final behaviour, and we may regard our perturbation above which is not chosen to coincide with the unstable modes, but just overlap them, as giving this generic evolution. It would be interesting to explore whether more exotic behaviours could be found in the case of large charges where there are multiple unstable modes which may be tuned initially in the seed perturbation. 

This similar qualitative behaviour to the uncharged case is perhaps to be expected. Our electric EM flow preserves surface gravity and electric potential difference to the horizon. At fixed electric potential and surface gravity there is only one infilling RN solution, which is precisely the initially perturbed fixed point. Thus, as in the uncharged case, there is no natural end state solution to flow to, except flat spacetime (with a constant electric potential). While we haven't explored this in detail here, one may resolve the singularity formed by the shrinking horizon case, and presumably this then flows to flat spacetime after the topology is suitably changed, as happens in the uncharged case. We may understand the other growing horizon as the flow `searching' for a larger stable black hole. If one puts the flow in a box, then as in the uncharged case, a new large black hole solution exists~\cite{Braden:1990hw}, and this expanding horizon flow would presumably settle down to that solution.

\begin{figure}
\centerline{
  \includegraphics[width=8cm]{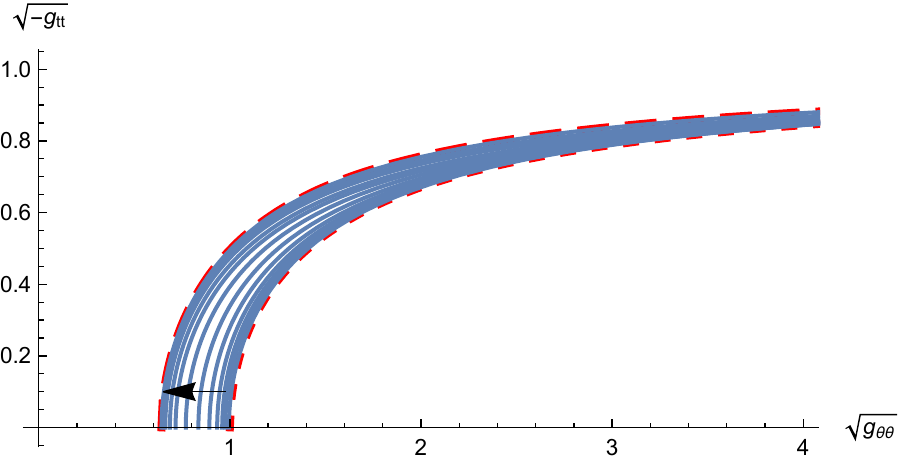}   \includegraphics[width=8cm]{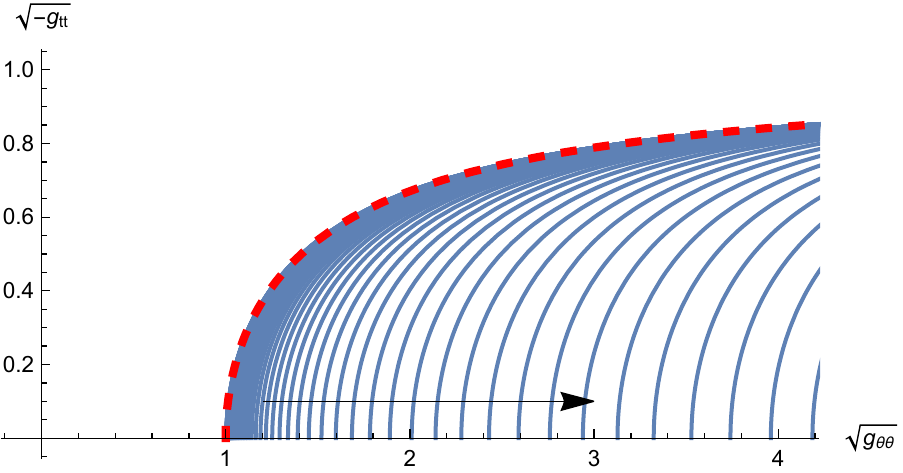} 
}
\centerline{
  \includegraphics[width=8cm]{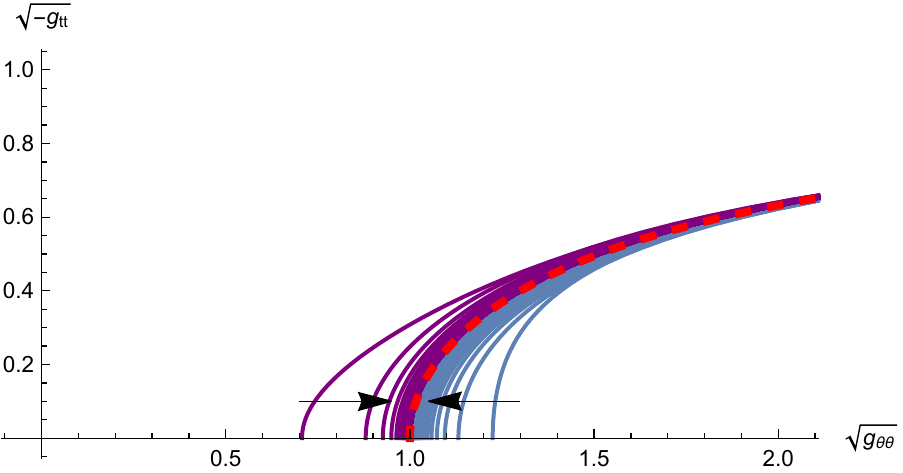} 
  }
  \caption{\label{fig:NonExtMag}
  Figure showing curves of $\sqrt{-g_{tt}}$ against $\sqrt{g_{\theta\theta}}$ at constant intervals now for the magnetic EM flow.
  The top left and right hand frames are similar to those of the previous figure, being for an unstable magnetic RN solution with $r_- = 0.2$ perturbed initially to have smaller horizon size (left) or larger horizon size (right). The latter gives the same expanding horizon behaviour as that of the uncharged or electric cases -- here $\Delta \lambda = 0.5$. The red dashed curve again shows the geometry of the initial RN fixed point that is perturbed. In the initially contracting case here a new stable magnetic RN solution exists for the same surface gravity and charge, and the flow asymptotes to this -- here it has $r_+ = 0.64$ and we show the new stable fixed point geometry using longer red dashes. The flow curves in this case are plotted for $\lambda = 0, 10, 12, \ldots, 22, 24, 30, 50$ and $1000$. The lower frame shows flows from a  non-linear deformation of a fixed point with larger charge $r_+ = 0.4$, so it is stable. Two flows are shown, one with larger horizon initially (blue curves) and one with smaller  (purple curves). Both quickly return to the stable fixed point (shown by the red dashed curve) under the flow (for both curves are shown at intervals $\Delta \lambda = 0.5$). Arrows indicate the sense of change of the geometry under increasing flow time.
    }
\end{figure}

Now let us discuss the case of the magnetic flow, and recall that the magnetic charge is preserved by the flow. Analogous plots of $\sqrt{-g_{tt}}$ against $\sqrt{g_{\theta\theta}}$ are shown in figure~\ref{fig:NonExtMag} for two situations, one below the critical charge and one above. Above the critical charge our linear analysis has shown no negative modes, and indeed we see that large initial deformations of the RN solution flow back to it. The figure shows the example of $r_- = 0.4$ (recall that $r_- > 1/3$ gives charges larger than the critical one) for an initially large deformation,
\be
T = 1 - (1 - \rho^2) r_- \; , \quad A = 1/T \; , \quad S = 1 \pm 0.5 \times (1 - \rho^2)^4 \; .
\ee
The plot shows both the positive and negative initial deformations on the same axes, and both flow back to the RN fixed point. We observe this same behaviour for all charges greater than the critical one, so $r_- > 1/3$, and for a variety of non-linear initial deformations to the geometry. 

The case of sub-extremal charge, so $r_- < 1/3$, is much more interesting. In this case we expect from~\cite{Monteiro:2008wr} one unstable mode of the flow, and this is confirmed by our linear analysis. There is now an important difference from the electric case. While there, at fixed surface gravity and electric potential there is one solution, the perturbed fixed point, now in the magnetic case where magnetic charge is conserved, one has an additional fixed point. This has the same surface gravity and charge as the unstable RN fixed point with $r_+ = 1$, this second solution now having $r_+ < 1$ and being a stable fixed point. Thus for an initial perturbation that reduces the horizon size, rather than flowing to zero size and a singularity, another option is to flow to this second stable fixed point with its smaller horizon, but the same charge. This is precisely what we find occurs in our simulations. For an initial perturbation similar to that in the electric case (although now there is no potential), so,
\be
\label{eq:nonextmagpert}
T = 1 - (1 - \rho^2) r_- \; , \quad A = 1/T \; , \quad S = 1 \pm 0.01 \times (1 - \rho^2)^4 
\ee
we observe that the flow with the initial RN deformed to have a smaller horizon flows to the second RN fixed point. This is shown in the figure~\ref{fig:NonExtMag} for the case $r_- = 0.2$, but we see the same behaviour for all initial charges $r_- < 1/3$ that we have studied. The initial unstable fixed point has horizon size governed by having $r_+ = 1$, and the stable end point then has $r_-$ given by the other (positive) root of the cubic,
\be
\label{eq:nonextmagstablesoln}
(1- r_-) r_+^3 - r_+^2 + r_- = 0 \; .
\ee
Unlike the uncharged or electric cases, where with a change of topology that removes the horizon we may have an infilling flat spacetime fixed point (with constant potential in the electric case), in this magnetic setting where the magnetic charge is fixed by the asymptotic boundary conditions, this is not possible. With a flat spacetime topology there is no way to infill the geometry smoothly to a give a fixed point, and yet carry the magnetic charge. 
Thus it intuitively makes sense that in this magnetic case we see the horizon shrinking flow go to the stable charged black hole, rather than choose to shrink the horizon to zero size and a singularity, which whilst it could be resolved, could not now flow to a flat spacetime fixed point due to the flow preserving magnetic charge. 
Another way to understand this different behaviour is that the charge enclosed at any radius is the same, as there is no charged matter, so the horizon carries the fixed magnetic charge. When the horizon shrinks, the backreaction from the fixed charge becomes larger, and its contribution to the stress tensor would diverge if the horizon shrinks to zero size. Thus this trapped magnetic flux acts to prevent the horizon shrinking.

As we see in the figure~\ref{fig:NonExtMag}, the flow that initially increases the horizon size of the unstable fixed point, appears to continue to do so in an accelerated manner as in the uncharged or electric cases. We see this same behaviour for all $r_- < 1/3$ as we perturb the unstable RN fixed point in a manner initially expanding the horizon. As mentioned above for the electric case, presumably if we placed this system in a spherical box, a new stable `large' black hole fixed point would exist, and this horizon expanding flow would asymptote to this. In the asymptotically flat setting we consider here, the flow expands the horizon forever, always searching, but not finding, a large stable black hole to settle on.

\begin{figure}
\centerline{
  \includegraphics[width=12cm]{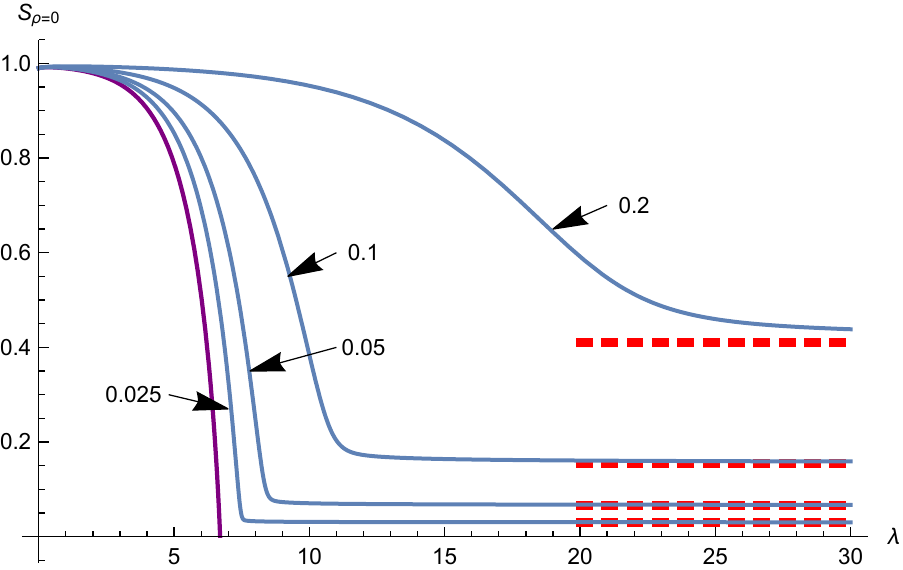} 
  }
  \caption{\label{fig:NonExtMagSCurves}
  Figure showing curves of $S$ evaluated at the horizon, $\rho = 0$, against flow time for magnetic flows of perturbed magnetic RN which initially decrease the horizon size. This shows how the (square of the) horizon radius flows. The purple curve is the uncharged case, where the horizon shrinks to a singularity at finite flow time, $\lambda \sim 6.7$. The other curves show charges given by $r_- = 0.025, 0.05, 0.1$ and $0.2$. We see that for all these charges the horizon asymptotes to the stable one for that charge and surface gravity (the values of $S$ for the stable black hole are shown as red dashed lines and are computed from equation~\eqref{eq:nonextmagstablesoln}). As the charge is decreased, the curves become more similar to the uncharged case until the point that the horizon settles on that of the smaller stable black hole solution.
    }
\end{figure}

Finally we show in figure~\ref{fig:NonExtMagSCurves} that as the magnetic charge is decreased to zero, we recover the shrinking flow to a zero sized singularity at the horizon. We plot the metric function $S$ at the horizon, $\rho = 0$, as a function of flow time, which gives the squared radius of the horizon sphere. The plot shows the horizon flow for the same initial metric perturbation as in~\eqref{eq:nonextmagpert} with various charges, including the zero charge case. As discussed in~\cite{Headrick:2006ti}, for zero charge this develops a singularity, here at approximately $\lambda \sim 6.7$ where the horizon shrinks to zero radius. The largest charge shown is that where $r_- = 0.2$ as in figure~\ref{fig:NonExtMag} and we see the horizon settles down to the radius of the corresponding stable black hole with that same charge and surface gravity (so $r_+ = 0.64$). We see that for smaller charges the flow of the black holes approach the correspondingly smaller stable black hole solutions determined by~\eqref{eq:nonextmagstablesoln}. The curve of $S$ against flow time tends to that of the uncharged case, up to the point where the horizon stops shrinking and settles on the smaller stable horizon.

\section{Spherically symmetric flows about extremal Reissner-Nordstr\"om}
\label{sec:ExtRNflows}

For static extremal black hole spacetimes, with pure electric or magnetic charge, we have argued above that we may  perform static EM flows, and further that the flow of the near horizon geometry decouples from the exterior. 
Following our discussion of the stability of static EM flows in the spherically symmetric case about non-extremal RN, it is natural to extend this to consider the extremal case. To do this we consider the general spherically symmetric metric using the coordinates introduced earlier in equation~\eqref{eq:RNextmetric}, taking,
\be
\label{eq:extmetric}
ds^2 = - \rho'^2 T  dt^2 +  r(\rho')^4 A \frac{d\rho'^2}{\rho'^2}  + r(\rho')^2 S  d\Omega^2 \; .
\ee
From our earlier discussion, in order to have a smooth extremal horizon located at $\rho' = 0$ we should have  $T, A, S$ being smooth functions of $\rho'$ and further satisfying,
\be
\label{eq:extsphsmooth}
\left. \left( T - A \right) \right|_{\rho' = 0} = 0 \; , \quad \left. \frac{\partial_{\rho'} \left( T - A \right)}{T} \right|_{\rho' = 0} - 4 = \psi'
\ee
for some constant $\psi'$.
Now considering the static EM flow, $T, A, S$ become also functions of flow time $\lambda$. For the pure magnetic case, we simply have,
\be
\label{eq:magans}
A_{mag} = - \cos{\theta} d\phi
\ee
as this is preserved by the flow. In the electric case, instead, we take,
\be
\label{eq:elecans}
A_{elec} = \rho' \Phi dt
\ee
which is smooth provided $\Phi$ is a smooth function of $\rho'$ and flow time. One may check explicitly that the spherically symmetric static EM flow equations preserve the smoothness conditions~\eqref{eq:extsphsmooth}, following our general analysis earlier. From our discussion in section~\ref{sec:setup}, using our choice of units, the extremal RN fixed point is given by $T = A = S = 1$, with $\Phi = 1$ in the electric case, so that $\psi' = -4$.

Firstly we may explicitly solve the near horizon non-linear flow equations in both the electric and magnetic cases.
Specializing our general discussion in section~\ref{sec:extremal} to our spherically symmetric ansatz, the near horizon form of the metric is,
\be
\label{eq:nhform}
ds^2_{NH} = - T_0  \left( \rho'^2  dt^2 +  \frac{d\rho'^2}{\rho'^2} \right) + S_0  d\Omega^2 
\ee
and in the electric case,
\be
\label{eq:extelecans}
A^{NH}_{elec} = \rho' \Phi_0 dt
\ee
with $T_0$, $S_0$ being functions of only flow time. As discussed in section~\ref{sec:extremal}, for the electric near horizon flow the potential $\Phi_0$ is a constant in flow time. 
As one might expect from the non-extremal case above, solving this near horizon flow we will find the magnetic RN solution is a stable fixed point of the magnetic flow, but the electric solution is unstable. 

In either the magnetic or electric case we may fix the near horizon metric to be that of RN, and then consider the flow in the exterior. We will simulate these flows numerically and fully non-linearly for flow parameter $\flows = 1$. In the magnetic case the flows approach magnetic RN at late times. However in the electric case, even with the horizon fixed to be that of extremal RN, we see generic initial data -- even that corresponding to small initial perturbations -- seem to develop to a singularity. 

\subsection{Near horizon Electric flow}

The electrostatic near horizon flow is determined by the flow equations,
\be\label{enhf}
T_0'(\lambda) = 2   - 2 \frac{\Phi_0^2}{T_0}    \; , \quad S_0'(\lambda) = - 2 + \frac{2 \Phi_0^2 S_0(\lambda)}{T_0(\lambda)^2} \; .
\label{eq:horizonDE}
\ee
The RN near horizon fixed point is given by $T_0(\lambda) = S_0(\lambda) = \Phi_0$, and we have chosen units $\Phi_0 = 1$. 
We may expand about this fixed point perturbatively and solve \eqref{eq:horizonDE} giving,
\be\label{param}
\begin{split}
   T_0(\lambda) &= 1 + \epsilon  e^{2 \lambda} a  \; , \quad S_0(\lambda) =  1 + \epsilon  e^{2 \lambda} \left( b - 4 \lambda a \right)  \; , \quad \Phi_0 = 1
\end{split}
\ee
to first order in $\epsilon$, where $a$ and $b$ are integration constants. Note that we do not perturb $\Phi_0$ as it is constant in flow time, and a perturbation would simply correspond to changing the charge of the fixed point solution, which by a change of units can be mapped back to one.
We thus see the fixed point is unstable, and there are two relevant deformations of it, parameterized by $\epsilon a$ and $\epsilon b$.
Starting with the RN near horizon solution as $\lambda \to - \infty$, adding these relevant deformations will then flow the near horizon metric away from near horizon RN for finite $\lambda$.

The simplest class of non-linear solutions which tend to the RN solution for $\lambda \to - \infty$, is that corresponding to $T_0$ being constant, $T_0(\lambda) = 1$. The flow equation for $S_0$ consequently linearizes, so that the non-linear solution for $S_0$ is simply,
\be
\label{eq:backreactedNH}
S_0(\lambda) = 1 +  \epsilon \, e^{2 \lambda} \; .
\ee
for a constant $\epsilon$.
We see that for $\epsilon < 0$ the horizon sphere shrinks to zero size at finite flow time $\lambda$. Alternatively for $\epsilon > 0$ it expands forever. 
We thus see that the extremal RN black hole is an unstable fixed point of the RN flow, with behaviour reminiscient of the non-extremal Schwarzschild black hole under Ricci flow~\cite{Headrick:2006ti}.

The general near horizon solution has non-constant $T_0$, and is given by,
\be\label{toexp}
T_0(\lambda) & = &  \left(1 + P_\pm( \lambda ) \right) \nl
S_0(\lambda) & = &  \frac{\left( 1 + \left( c - 4(\lambda-\lambda_0) \right) P_\pm(\lambda) + P_\pm(\lambda)^2 \right)}{1 + P_\pm(\lambda)}
\ee
where we have defined,
\be
P_\pm(\lambda) =  W_0\left( \pm e^{2 ( \lambda - \lambda_0 )} \right)
\ee
with $W_0(x)$, the Lambert W function that gives the principle solution of $y$ for the relation $x = y e^y$. The constant $\lambda_0$ arises from translation invariance of the equations in $\lambda$, and $c$ is the remaining constant of integration that determines the different behaviours.
Asymptotically as $\lambda \to - \infty$ we have,
\be\label{Wdef}
P_\pm(\lambda) \to \pm e^{2 ( \lambda - \lambda_0 )}
\ee
and the function $W_0(x)$ exists for all $x > -1/e$ and is monotonically increasing. At $x = -1/e$ it takes the value minus one. 



\begin{figure}[th]
\includegraphics[width=12cm]{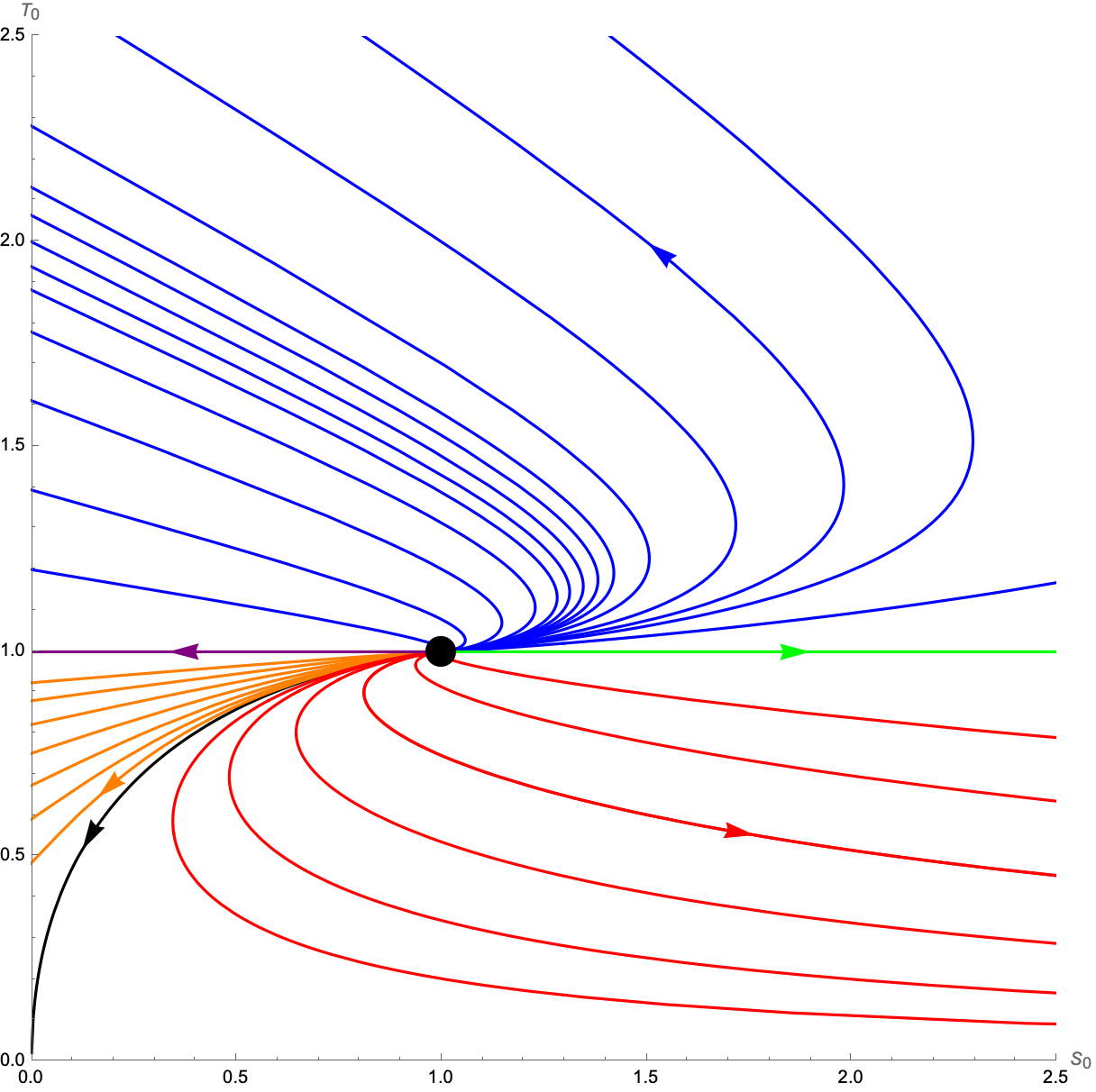}
\caption{\label{analyticnh} Analytic behaviour of the near horizon electric flow for small perturbations of the RN fixed point with $\Phi_0=1$. 
The various colors correspond to different choices for $c$ and the branches $P_{\pm}$ and are detailed in the main text.
The arrows point in the direction of the flow, while the black dot highlights the fixed point. The near horizon magnetic flow can be obtained by reversing the arrows and swapping $S_0$ and $T_0$.
  }
\end{figure}

The analytic behaviour of the near horizon flow is captured in Figure \eqref{analyticnh}. The sign in the definition \eqref{Wdef} of $P_\pm\left(\lambda\right)$ defines two distinct branches for evolutions  away from the fixed point. For these, the various trajectories in the $\left(S_0,T_0\right)$ plane are then labelled by the value of $c$.
For the  $P_+\left(\lambda\right)$ branch (shown as blue curves in the figure), then $S_0 \to 0$ in finite flow time with $T_0$ remaining finite. For the  $P_-\left(\lambda\right)$ branch with $c > 0$ (shown as orange curves) the same behaviour occurs. However, taking instead $c < 0$ (shown as red curves) then $S_0 \to \infty$ and $T_0 \to 0$ in a finite flow time. In the special case $c = 0$ for the  $P_-\left(\lambda\right)$ branch (shown as the black curve) which separates these two behaviours, both $T_0$ and $S_0$ go to zero at the same finite flow time. Finally taking the limit $c \to \pm \infty$ for the $P_\pm\left(\lambda\right)$ branches results in the two $T_0 = 1$ behaviours in~\eqref{eq:backreactedNH} above, one with $S_0$ that shrinks to zero in finite flow time (purple in the figure) and the other with growing $S_0$ (green in the figure) so that the horizon expands forever. The only behaviour which exists for all flow times is this last one.


The Ricci scalar of the near horizon geometry, which is the Ricci scalar of the full spacetime restricted to $\rho = 0$, is simply given by,
\be
\left. R \right|_{\rho=0} = - \frac{2}{T_0} + \frac{2}{S_0}
\ee
For all finite values of $c$ then either $S_0 \to 0$ at a finite flow time, with finite $T_0$, or alternatively $T_0 \to 0$ with $S_0 \to \infty$ at a finite flow time. In both cases the Ricci scalar diverges, showing that the geometry encounters a curvature singularity at finite flow time. The special case $c = 0$ where $T_0 \to 0$ and $S_0 \to 0$ could allow a cancellation between the two terms in the Ricci scalar, but calculation reveals it is still singular.
The only exception to a finite flow time singularity is the flow $T_0 = 1$ with expanding horizon, where the flow exists for all time, and the scalar curvature tends to $\left. R \right|_{r=0} \to - 2/T_0$.
We note that any solution to the Einstein-Maxwell equations has vanishing Ricci scalar (since the stress tensor is traceless), and hence the asymptotic geometry that this flow tends to cannot be the near horizon geometry to an Einstein-Maxwell solution due to its non-vanishing Ricci scalar. 


\subsection{Near horizon Magnetic flow}
Now we consider the magnetic case where the near horizon flow equations are,
\be\label{mfnh}
T_0'(\lambda) = 2   - 2 \frac{T_0}{S_0^2}    \; , \quad S_0'(\lambda) = - 2 + \frac{2}{S_0(\lambda)} \; .
\label{eq:horizonDEM}
\ee
Then introducing functions $\Bar{T}_0\left(\lambda\right)\equiv S_0\left(-\lambda\right)$ and $\Bar{S}_0\left(\lambda\right)\equiv T_0\left(-\lambda\right)$, we derive the flow equations for these,
\be\label{flippedeq}
\Bar{T}_0'\left(\lambda\right)=2 - \frac{2}{\Bar{T}_0} \; ,\quad \Bar{S}_0'\left(\lambda\right)=-2   + 2 \frac{\Bar{S}_0}{\Bar{T}_0^2} 
\ee
which are precisely the near horizon electric equations~\eqref{enhf} taking $S_0 \to \bar{S}_0$ and $T_0 \to \bar{T}_0$. Hence, the near horizon magnetic flow is equivalent to the electric one, after exchanging $T_0$ and $S_0$, and flipping the sign of the flow time $\lambda$.
This implies that the unstable near horizon RN fixed point that lies in the infinite past of the electric near horizon flow translates into a non-linearly stable fixed point of the magnetic near horizon flow.

\subsection{Flows of the full extremal spacetimes}

Finally we numerically investigate the non-linear spherically symmetric flows starting with a deformation of the extremal electric and magnetic RN solution that preserves extremality. Again we consider only the case $\flows = 1$. We solve the full non-linear flow for the metric functions $T, A, S$, and in the electric case $\Phi$ too, discretizing in the spacetime coordinates using sixth order finite differencing as for the non-extremal case discussed earlier. We employ a second order accurate Crank-Nicholson differencing in flow time, and solve the implicit system at each time step using an iterative solver. The code correctly reproduces the continuum limit, and the results we present here are for $N = 200$ spatial points, and for a flow time step of $5 \times 10^{-5}$. In the non-extremal setting we needed a boundary condition at the horzion for our metric functions. However in the extremal case the flow equations are simply imposed at the horizon, and as we have discussed above, decouple from the exterior flow as they involve no radial derivatives. While we have the analytic solution for any initial data for the near horizon flow, it is convenient here to simply evolve the near horizon flow numerically. We then check it correctly reproduces the analytic solutions.

\begin{figure}
\centerline{
  \includegraphics[width=14cm]{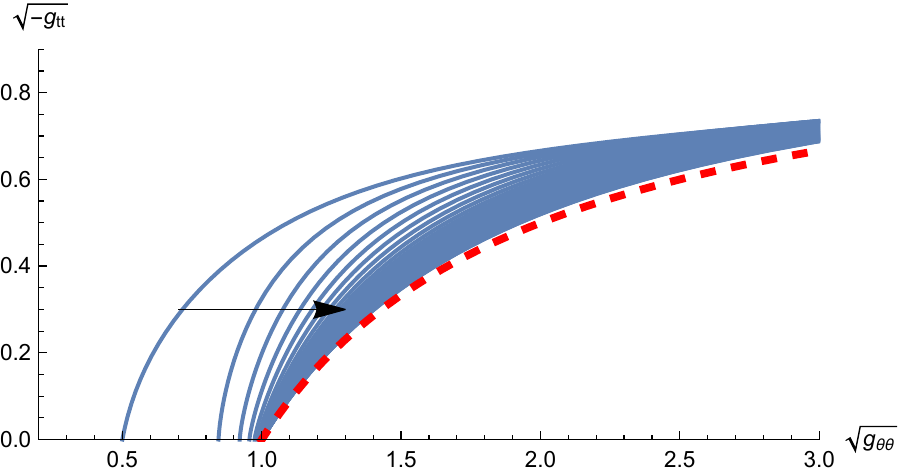} 
}
  \caption{\label{fig:Mag}
  Figure showing curves of $\sqrt{-g_{tt}}$ plotted against $\sqrt{g_{\theta\theta}}$ at constant intervals ($\Delta \lambda = 0.25$) of flow time, for a non-linear initial deformation of the extremal magnetic RN solution. The RN fixed point is shown in dashed red, and we see the initally deformed geometry asymptotes back towards extremal RN as the flow proceeds.
    }
\end{figure}

Our expectation is that deformations of the magnetic RN solution should return to it under the flow, and indeed this is what we see. An example is shown in figure~\ref{fig:Mag} where we start with a non-linear deformation of the magnetic solution, initially deforming the metric functions as,
\be
T = A =  1 + 2 (1 - \rho'^2)^4 \; , \quad S = 1 - \frac{3}{4} (1 - \rho'^2)^4   \; .
\ee
We note that these are compatible with smoothness of the horizon, as in equation~\eqref{eq:extsphsmooth}, and preserve $\psi' = -4$. We already know that the near horizon flow, ie. the behaviour at $\rho' = 0$, must return to the magnetic near horizon RN geometry. In figure~\ref{fig:Mag} we display the geometric invariant $\sqrt{-g_{tt}}$ plotted against $\sqrt{g_{\theta\theta}}$. These curves are shown at intervals of flow time $\Delta \lambda = 0.25$, and we see they asymptote back towards the magnetic RN fixed point. This behaviour is generic for other non-linear deformations we have implemented, provided they are not too large (when they can potentially give rise to singularities away from the horizon). Thus at least in the spherically symmetric setting, the extremal magnetic RN solution appears to be a stable fixed point of the non-linear flow for the full extremal horizon spacetime.

\begin{figure}
\centerline{
  \includegraphics[width=8cm]{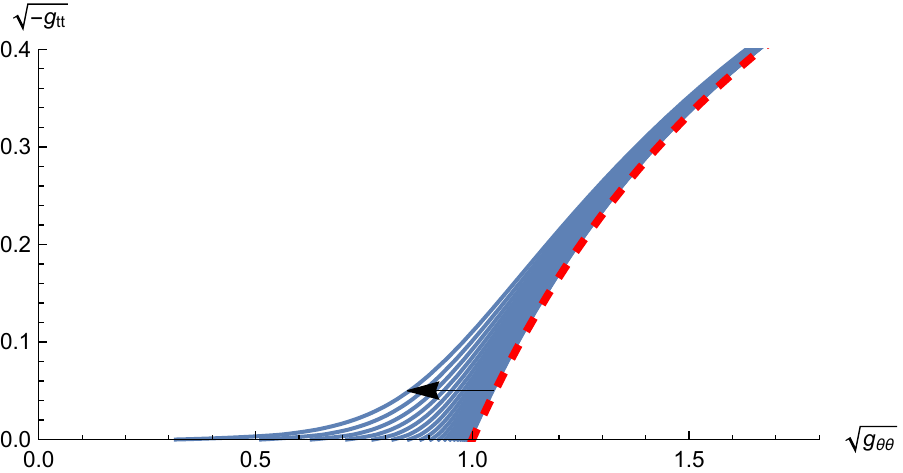}   \includegraphics[width=8cm]{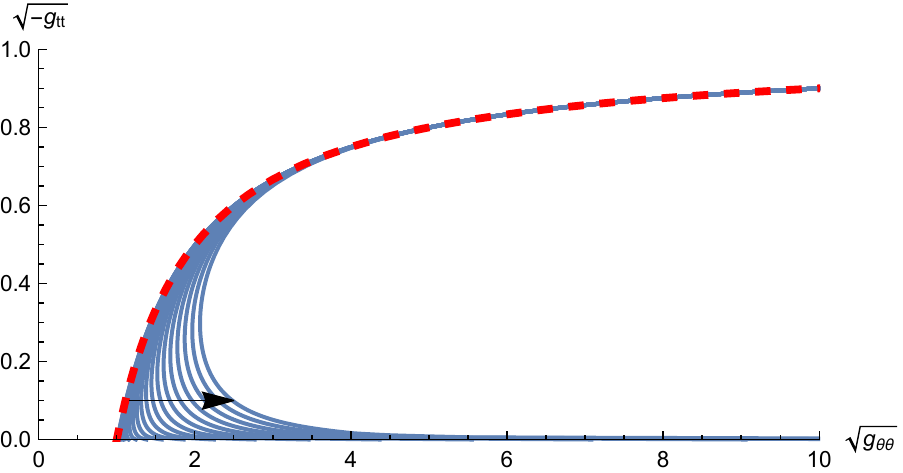} 
}
  \caption{\label{fig:Elec1}
  Similar figure to the previous one, but now for flows that begin with a small perturbation of the electrically charged extremal RN solution. The perturbation is subtracted and added to initially decrease and increase the horizon size, giving the left and right frames plotted for flow time intervals $\Delta\lambda = 0.1$ and $0.25$ respectively. 
  The flow behaviour at the extremal horizon decouples and follows the near horizon flow, leading to a collapse in finite flow time or exponential horizon expansion. We see the exterior of the geometry follows this near horzion behaviour in a smooth manner. In the collapsing case the singularity occurs at $\lambda = 2.30$. In the expanding case curves are shown for flow times up to $\lambda = 5$, when gradients near the horizon become difficult to numerically resolve.
    }
\end{figure}

Now let us consider the electrically charged RN fixed point. Our near horizon analysis already indicates it is unstable, and in figure~\ref{fig:Elec1} we follow two flows where the fixed point is initially perturbed by taking,
\be
S = 1 \pm 0.01 \times (1 - \rho'^2)^2
\ee
with $T = A = \Phi = 1$. From our near horizon analysis for the negative sign above we know the horizon behaviour is then $T = 1$ and $S = 1 - 0.01 e^{2 \lambda}$ at $\rho' = 0$, and so the horizon sphere will shrink to zero size at time $\lambda = 2.30$.
For the positive case the horizon behaviour is $T = 1$ and $S = 1 + 0.01 e^{2 \lambda}$ and so the horizon sphere exponentially expands in flow time. Simulating the full flows we again plot the invariant $\sqrt{-g_{tt}}$ against $\sqrt{g_{\theta\theta}}$ for both cases in the figure~\ref{fig:Elec1}. We indeed see these behaviours at the horizon, with the geometry away from the horizon responding but remaining smooth.  In the expanding case the simulation is run until flow time $\lambda \simeq 5$ when the function $S$ becomes too large ($S \sim O(10^2)$) at the horizon to properly resolve gradients there.
In the case where the sphere shrinks to zero size we expect that one can resolve the singularity, as done in~\cite{Headrick:2006ti}, to continue the flow through to flat spacetime.

\begin{figure}
\centerline{
  \includegraphics[width=8cm]{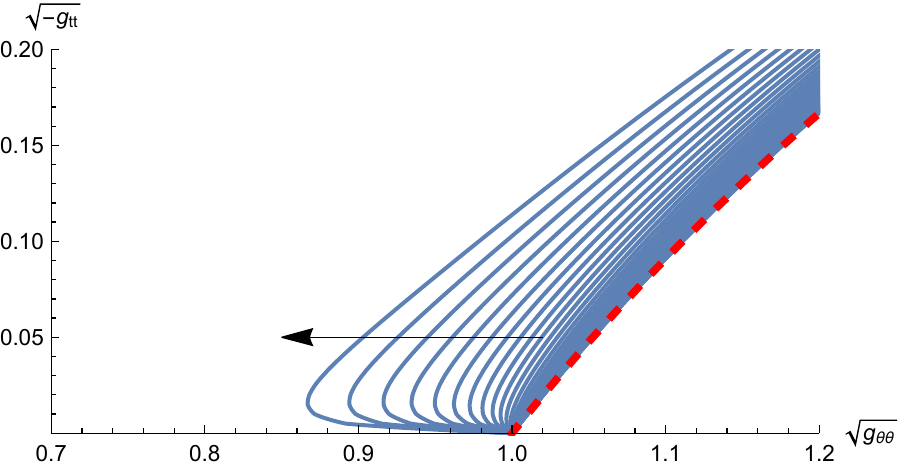}  \includegraphics[width=8cm]{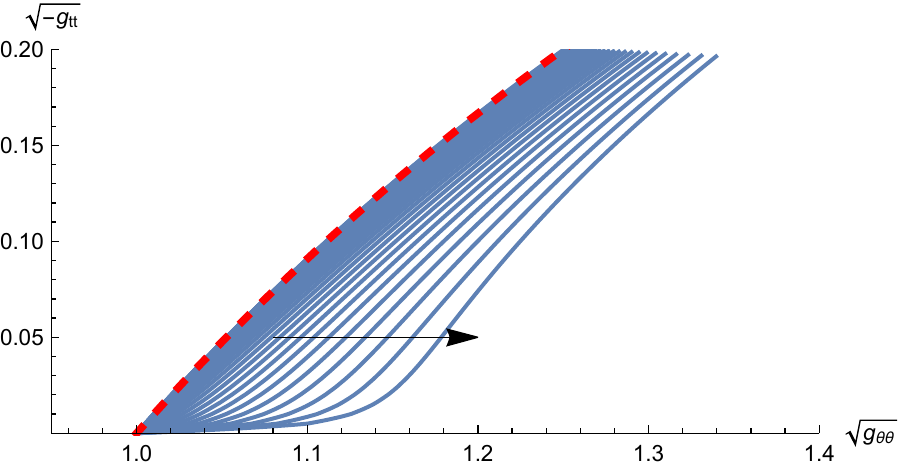} 
}
\centerline{
  \includegraphics[width=8cm]{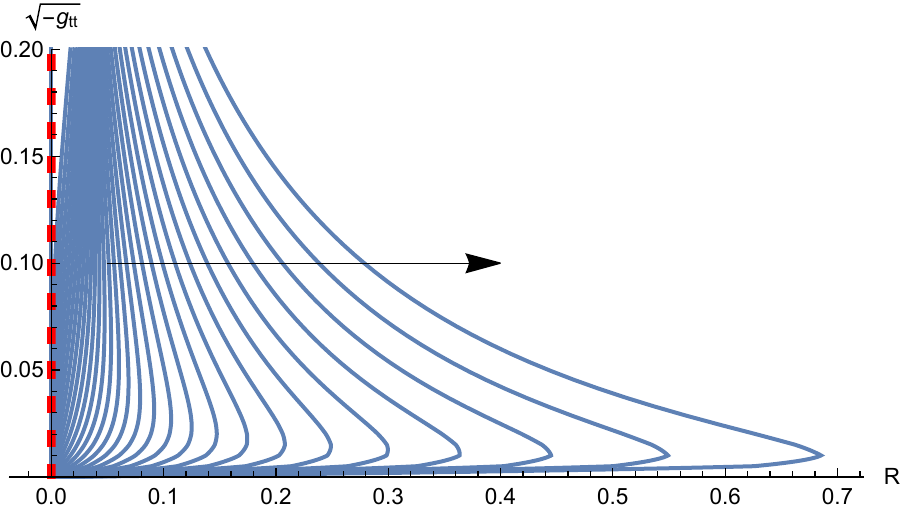}  \includegraphics[width=8cm]{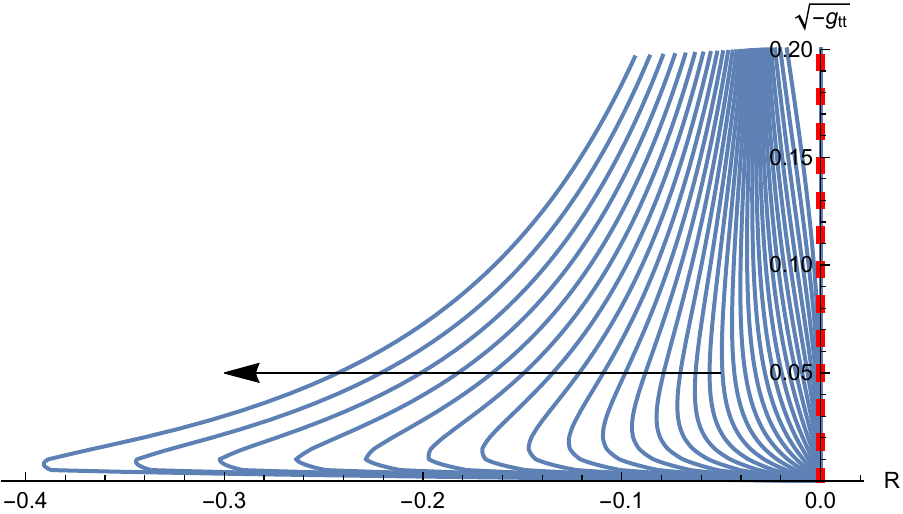} 
}
  \caption{\label{fig:Elec2}
  The top panels are similar to the previous figure, but now the two flows depicted are for an initial small perturbation of extremal electrically charged RN that preserve the horizon geometry so that the near horizon geometry remains that of extremal RN throughout the flow. The same perturbation is both added and subtracted to RN, generating the top left and right panels. In both cases the flow time interval between the curves is $\Delta \lambda = 0.1$.
  While the horizon geometry is fixed, we still see extremal RN is unstable as the perturbation grows in the exterior. This appears to build up curvature near, but not at, the horizon. This is confirmed in the bottom two panels, where the Ricci scalar is plotted against $\sqrt{-g_{tt}}$. This is zero for the RN solutions, and is fixed to zero at the horizon as the near horizon geometry is unperturbed here, but we see it appears to grow in the vicinity of the horizon. The curves are plotted up to a flow time $\lambda = 3$ when gradients become too large to accurately resolve. While it appears a singularity forms, it remains unclear whether this occurs in a finite flow time.
    }
\end{figure}

Finally we can consider the case where we do not initially perturb the horizon geometry. We show here the example deformation,
\be
S = 1 \pm 0.5\times \rho'^4 (1 - \rho'^2)^2
\ee
with $T = A = \Phi = 1$, and again note this preserves smoothness of the extremal horizon with $\psi' = -4$. The near horizon analysis then implies $T = S = 1$ for all flow time at the horizon $\rho' =0$. The plethora of unstable modes of the non-extremal flow suggests that while the horizon geometry is pinned to that of the extremal electric RN, the exterior spacetime may be unstable. Indeed this appears to be the case. In figure~\ref{fig:Elec2} we plot the same invariants as in the previous figures for both these flows, now zooming in on the region near the horizon. The horizon size indeed remains fixed, and the flow away from the horizon remains smooth for some time. We show the flow at time steps $\Delta \lambda = 0.1$ up to a flow time of $\lambda = 3$, when in both cases the gradients near the horizon become very large. In the same figure we show the Ricci scalar, $R$, plotted against $\sqrt{-g_{tt}}$ near the horizon and see that it becomes large and localized just outside the horizon -- recall that for RN the Ricci scalar $R = 0$, and for these flows it is fixed to have its RN value at the horizon as the near horizon geometry is pinned to that of RN. The maximum absolute value of the Ricci scalar appears to grow in an unbounded way as the flow proceeds. The maximum also becomes nearer to the horizon, making it difficult to numerically resolve at late flow times. 
We see the same behaviour for other perturbations we have tested for the electric extremal RN solution where the near horizon geometry is not deformed. 

These results appear to confirm that the extremal electric RN solution is unstable to deformations preserving extremality. Deformations that perturb the horizon obviously flow away, governed by the unstable near horizon flows discussed earlier. More interestingly, if the near horizon geometry is not perturbed, and thus stays that of extremal RN, the exterior solution then appears to flow away from RN, developing increasing curvature that is increasingly localized in the vicinity of the horizon as the flow proceeds.

\section{Summary}

We have explored the natural geometric flow for Einstein gravity coupled to a Maxwell field in the asymptotically flat setting. We have shown that for static geometries and Maxwell fields this flow is well-posed taking either purely electric or magnetic potentials, when it is parabolic in character. We have also shown that it preserves smooth non-extremal and extremal horizons. The surface gravity of a horizon is preserved by the flow. In the electric case the potential difference between infinity and the horizon is fixed by the natural asymptotic boundary conditions of fixing the gauge potential there. In the magnetic case, the natural boundary condition of fixing the asymptotic magnetic potential results in a conserved magnetic charge under the flow. Thus we may regard the electric and magnetic flows as being conjugate to each other in the sense that the electric one preserves the potential, while the magnetic flow preserves charge.

We have also discussed how the electric flow naively preserves charge, in the sense that at finite flow time, the charge computed asymptotically will remain unchanged. This is similar to how ADM mass is preserved under Ricci flow for finite times, as shown in~\cite{Oliynyk:2006nr}. However we have argued that this is not a physical conservation of charge, and the charge computed within a very large but finite radius will change in flow time. It simply takes an infinite time for this change of charge to be communicated to spatial infinity, given the diffusive character of the flow. Thus physically while the electric potential is fixed by boundary conditions of the electric EM flow, we should regard the electric charge as changing under the flow.

The fixed points of these flows are the RN solutions. In the non-extremal case the negative mode analysis of Euclidean magnetic RN suggests we should expect the magnetric RN fixed point to be unstable under the magnetic EM flow for charges below a critical value and stable above, with the critical charge being related to that where the specific heat at constant charge changes sign. By studying linear instabilities of the RN solution numerically for spherical symmetry we confirm this expectation. For the electric EM flow we find the electric RN fixed points are always unstable, and our numerical analysis of the instabilities, perhaps surprisingly, shows an increasing number of unstable modes as extremal charge is approached. While in the electric case there is no obvious direct connection to Euclidean negative modes and  thermodynamic stability, we note that this flow instability is qualitiatively similar to the thermodynamic stability, where electric RN is always unstable at fixed temperature and potential. It is in fact natural to compare it with this ensemble as our electrostatic EM flow preserves the surface gravity and the electric potential difference.

It is also interesting to note the similar qualitative change that adding charge has to the case of the Gregory-Laflamme instability~\cite{Gregory:1993vy}. There in the charged setting it was shown in~\cite{Frolov:2009jr} that magnetic charge acts to stabilize a black string, whereas electric charge destabilizes it. That is precisely the qualitative character that charge has in our flows -- magnetic RN fixed points are more stable than their uncharged counter-parts, whereas electric RN appears to be more unstable, even having additional unstable modes as the charge is increased relative to horizon size.

The linear instabilities trigger flows from the fixed points. We numerically determine this behaviour by evolving the non-linear flows generated by initial deformations of the unstable RN fixed points in spherical symmetry.  In the uncharged case of Schwarzschild under Ricci flow, the single unstable mode either shrinks the black hole horizon to a singularity, which wants to be resolved to then flow to flat spacetime, or blows it up in a search for a large stable black hole (which only exists for a system in a box, and not in our asymptotically flat setting). We have found similar behaviour for generic initial perturbations to the unstable RN fixed point for the electric EM flow. The horizon shrinks to zero size in finite flow time or expands seemingly indefinitely. While we haven't explored this here, presumably one can resolve the shrinking flow singularity to then yield a flow to flat spacetime at constant potential, and likewise putting the system in a spherical box, have the expanding flow tend to a stable large black hole.
Thus in the electric case a similar behaviour to the uncharged case is observed.

The magnetic case is more interesting since starting with an unstable fixed point, there exists another infilling geometry with the same surface gravity and charge (recall these are conserved for the magnetic EM flow) that is stable and has smaller horizon. Indeed we see the flows seeded by a perturbation that decreases the unstable RN horizon initially does asymptote to this stable fixed point, rather than trying to flow to flat spacetime via a singularity -- here in the magnetic case, the fixed magnetic charge at the horizon acts to prevent it from shrinking to zero size. 

It is interesting that as for Ricci flow of uncharged Schwarschild, the non-extremal static electric and magnetic EM flow behaviour mirrors the expected real time evolution of these black holes in a bath at constant temperature. In this dynamical setting the thermodynamic instability is then related to a dynamical instability -- if a thermodynamically unstable black hole is in (unstable) equilibrium with a bath, we expect thermal fluctuations to initiate an unstable real time dynamics.

Having discussed these non-extremal flows, we then focussed on the extremal case. Here the near horizon flow decouples from that in the exterior and we have solved it analytically for spherical symmetry. We saw from the non-extremal analysis that the electric RN is unstable up to extremality. Indeed the near horizon flow shows this instability in the electric case, and we generically find finite flow time singularities resulting from the sphere shrinking to zero size or expanding forever. The magnetic case, as expected, is non-linearly stable for spherical symmetry.

While in the non-extremal case, the linear instabilties we have studied have had an exponential form in flow time, in the extremal electric case we clearly see from the analytic near horizon solution that instabilities may have a more complicated flow time dependence. In particular we see both exponential $e^{2 \lambda}$ dependence as well as a more complicated growing behaviour going as $\lambda e^{2\lambda}$ in the linear theory. It would be interesting to understand whether this non-exponential growth results from the proliferation of exponential instabilities as extremal charge is approached in the electric case.

For the extremal flows since the near horzion geometry decouples from the exterior, we may regard it as a boundary condition for this exterior flow. One might then imagine that if one perturbs the electric extremal RN fixed point so that the near horizon geometry is unchanged, one will only have stable flows. However our full numerical simulation of these non-linear spherical flows shows this is not the case. While deformations to the extremal magnetic RN that preserve extremality return to the stable fixed point for the magnetic flow, in the electric case any deformation apparently takes one away from the fixed point, even if the near horizon geometry is unperturbed. In particular in the electric case we find small extremal perturbations that preserve the near horizon geometry appear to generate an increasingly singular geometry outside the horizon, which is seen by the Ricci scalar growing, apparently without bound.

It would be very interesting to study and better understand this instability, in particular whether it generates a singularity in the vicinity of the horizon, and if so, whether this occurs in a finite flow time.
While at first glance it seems quite different in nature to the dynamical instability of extremal horzions due to Aretakis~\cite{Aretakis:2011ha,Aretakis:2011hc,Aretakis:2012ei,Lucietti:2012sf,Lucietti:2012xr}, it is very interesting that such an instability of extremal solutions also occurs in the geometric flow setting, and it would be interesting to explore whether there is any relation. In the non-extremal case, as discussed above, we may intuitively understand the flow outcomes as being qualitatively related to real time behaviour of the black holes in a thermal bath. It would be very interesting to understand if this singular behaviour in the extremal, zero temperature limit, also mirrors some real time behaviour, perhaps associated to quantum fluctuations.

Here we have focussed on the EM flow restricted to static spacetimes. However the DeTurck-Ricci flow remains parabolic for stationary spacetimes, and also preserves non-extremal horizon smoothness, despite the presence of ergoregions~\cite{Adam:2011dn,Wiseman:2011by}. We expect that the same will be true for the Einstein-Maxwell flow, and it would then be consistent to include both stationary electric and magnetic Maxwell fields in the same flow. While it seems reasonable that smooth stationary non-extremal horizons will be preserved, it is less obvious that this will hold in the extremal case, and it would be interesting to confirm this in future work.

Finally it is worth emphasizing that there is a constant flow parameter, $\flows$, which throughout our explicit calculations we have chosen to be one.  It gives the relative speed of the flow of the metric, compared to that of the Maxwell field, and it would certainly be interesting to understand what explicit flow solutions one obtains for different choices of $\flows$. Our initial exploration reveals a similar qualitative behaviour for the non-extremal and extremal flows for values close to one, such as $\tau = 0.5$ or $2$, but perhaps there are interesting transitions to other behaviours for certain values. 
Likewise it may be interesting to explore other choices of the superspace metric than $\mathrm{DeWitt}_{-1}$, so that the trace part of the metric is flowed at different speeds. 
Furthermore, one might extend the analysis to Einstein-Maxwell-dilaton theories, where in addition to the Maxwell field there will also be a scalar field to flow. If flow singularities were to occur, then studying the dilaton behaviour at these would allow for a natural connection to the Swampland distance conjecture. The addition of a dilaton field might also give natural flows that result from world-sheet RG flows.

\subsection*{Acknowledgments}
We are grateful to Stefanos Aretakis, James Lucietti and Martin Taylor for useful discussions.
This work was supported by STFC Consolidated Grant ST/T000791/1. The work of D.L. is supported by the Origins
Excellence Cluster and by the German-Israel-Project (DIP) on Holography and the Swampland.


\bibliographystyle{JHEP} 
\bibliography{flow}

\end{document}